\documentclass[a4paper,11pt]{article}

\usepackage{ifpdf}

\usepackage{epsfig,multicol,amsmath}
\usepackage[T1]{fontenc}

\usepackage{jheppub}
\usepackage{amsmath}     
\usepackage{epsfig,epsf}
\usepackage{graphicx}
\usepackage{rotating}
\usepackage{verbatim}
\def\beq{\begin{equation}}
\def\eeq{\end{equation}}
\def\bea{\begin{eqnarray}}
\def\eea{\end{eqnarray}}

\def\eq#1{{Eq.~(\ref{#1})}}
\def\fig#1{{Fig.~\ref{#1}}}

\newcommand{\bas}{\bar{\alpha}_S}
\newcommand{\ba}{\bar{\alpha}_{S 0}}
\newcommand{\as}{\alpha_S}

\newcommand{\Lb}{\left(}
\newcommand{\Rb}{\right)}
\renewcommand{\vec}[1]{\boldsymbol{#1}}
\newcommand{\dif}{\mathrm{d}}

\parskip=\itemsep               

\newcommand{\nn}{\nonumber}

\newcommand{\h}{\frac{1}{2}}

\newcommand{\al}{\alpha}

\newcommand{\pom}{I\!\!P}

\def\pom{{I\!\!P}}

\title{CGC/saturation approach: a new impact-parameter dependent model in the next-to-leading order of perturbative QCD.}
\author[a]{ Carlos Contreras,}
\author[a,b]{ Eugene ~ Levin,}
\author[c]{Rodrigo Meneses}
\author[a]{  and Irina Potashnikova}

\affiliation[a]{Departemento de F\'isica, Universidad T\'ecnica Federico Santa Mar\'ia, and Centro Cient\'ifico-\\
Tecnol\'ogico de Valpara\'iso, Avda. Espana 1680, Casilla 110-V, Valpara\'iso, Chile}
\affiliation[b]{Department of Particle Physics, School of Physics and Astronomy,
Raymond and Beverly Sackler
 Faculty of Exact Science, Tel Aviv University, Tel Aviv, 69978, Israel}
\affiliation[c]{Escuela de Ingenier\'\i a Civil, Facultad de Ingenier\'\i a, Universidad de Valpara\'\i so, Avda Errazuriz 1834, Valpara\'\i so, Chile}
\emailAdd{carlos.contreras@usm.cl}
\emailAdd{leving@post.tau.ac.il, eugeny.levin@usm.cl}
\emailAdd{rodrigo.meneses@uv.cl}
\emailAdd{irina.potashnikova@usm.cl}

\abstract{ This paper is the first attempt to build CGC/saturation model based on the next-to-leading order corrections to linear and non-linear evolution in QCD. We assume that the renormalization scale is the saturation momentum and  found that the scattering amplitude has geometric scaling behaviour deep in the saturation domain  with  the explicit formula of this behaviour at large $\tau = r^2 Q^2_s$. We built a model that include this behaviour, as well as the ingredients that has been known: (i) the behaviour of the scattering amplitude in the vicinity of the saturation momentum, using the NLO  BFKL kernel; (ii) the pre-asymptotic behaviour of $\ln\Lb Q^2_s\Lb Y \Rb\Rb$, as function of $Y$ and (iii)  the impact parameter behaviour of the saturation momentum,  which has exponential behaviour $\propto \exp\Lb -\, m\, b\Rb$ at large $b$.
We demonstrated that the model is able to describe the experimental data for the deep inelastic structure function. Despite this, our model has  difficulties that are  related to the small value of the QCD coupling at $Q_s\Lb Y_0\Rb$ and the large values of the saturation momentum, which indicate  the theoretical inconsistency of our  description.
}

\begin{document}

\maketitle
\flushbottom

\pagestyle{empty}

\mbox{}

\pagestyle{plain}

\setcounter{page}{1}


\section{Introduction.}

~~This paper is the  next step (see Ref. \cite{CLP}) in our attempt
to find an approach, based on Color Glass Condensate/saturation  effective theory  for high energy QCD (see  
Ref. \cite{KOLEB} for a review), which  includes the impact parameter dependance of the scattering amplitude.
Unfortunately, at the moment,  our efforts reduce to building a model which incorporates the main features of 
 the solution of the CGC/saturation equations, and  also contains a number of  phenomenological parameters for the non-perutbative QCD description of the large impact parameter dependance of the scattering amplitude.
 
 We are doomed to build models to introduce the main features of the CGC/saturation approach,                                   
since   the CGC/saturation equations do not   reproduce the correct behavior of the scattering amplitude at large impact parameters (see Ref. \cite{KW,FIIM}).   Such failure leads to the conclusion: we cannot trust the solution of the CGC/saturation equations, without the long distance non-perturbative corrections at large impact parameters. 

Indeed, for the scattering of a  dipole with size $r$,   with the nucleus, the CGC/saturation equations \cite{JIMWLK,BK}(see Eq.2.6 in Ref. \cite{KLT})  can be rewritten
for $N\Lb r, Y, Q_T=0\Rb = \int d^2 b\, N\Lb r, Y, b\Rb $  using the natural assumption that $r \,\ll \,R_A$ , where $R_A$ is the size of the nucleus. $N\Lb r, Y, Q_T=0\Rb $ is the infrared safe observable in perturbative QCD and, hence, we can expect that non-perturbative corrections for it, will be small.  The radius of the dipole increases with energy growth, but from high energy phenomenology we learned that this increase is of the order $\alpha'_\pom\, Y \ll R_A$  for $Y \leq 40$. Implicitly, we assume that the non-perturbative corrections change the power like increase with energy of the interaction radius, that follows from perturbative QCD  \cite{KW,FIIM},  to a logarithmic one, we believe that this change does not lead to  the violation of the  CGC/saturation equations.

However, for the interaction with a proton,  we do not even have  this, rather weak, argument and for a  hadron target we anticipate  large corrections to the CGC/saturation equations .  Real progress in theoretical understanding of confinement of quarks and gluon has not yet  been made , and  as a result, we do not know how to change the CGC/saturation equations to incorporate  confinement.  We have to build a  model which includes both   theoretical knowledge that stems from the CGC/saturation equations,  and  the phenomenological large $b$ behavior, which do not contradict   theoretical restrictions \cite{FROI,BRLE}.

Numerous attempts have been made over the past two decades (see Refs. \cite{CLP,SATMOD0,SATMOD1,BKL,SATMOD2,IIM,SATMOD3,SATMOD4,SATMOD5,SATMOD6,SATMOD7,SATMOD8,SATMOD9,SATMOD10, SATMOD11,SATMOD12,SATMOD13,SATMOD14,SATMOD15,SATMOD16,SATMOD17}) to build  such models. Therefore, we clarify, in the introduction,  the  aspects of our model  which are different.

The main difference of his paper from  others, is that we use the nonlinear Balitsky-Kovchegov(BK) equation  in the next-to-leading  order (NLO) of perturbative QCD,  that has been proven in Ref. \cite{NLOBK1,NLOBK2,NLOBK3}. The form of BK equation in the NLO shows that we can apply the method, suggested in Ref. \cite{LETU}, for determining the behavior of the solution to BK equation  deep inside  the saturation region. This behavior in the NLO is given in this paper.  It shows  geometric scaling behaviour as in the leading order of perturbative QCD,  for the renormalization scale which is equal to the saturation momentum $Q_s$.

                                                                                                                                                                                                                                                                                                                                                                                                                                                                                                                                                              We  only introduce the non-perturbative impact parameter behavior in the saturation momentum, accordingly to the  spirit of the  geometric scaling behavior of the scattering amplitude \cite{BALE,GS},  and to the semi-classical solution to the CGC/saturation equations   \cite{BKL}. Similar assumptions  for the non-perturbative $b$-behavior of the scattering amplitude, is typical  for most  models on the market (see Refs. \cite{SATMOD5,SATMOD6,SATMOD7,SATMOD8,SATMOD9,SATMOD12,SATMOD17}).
 In the choice of the $b$ behavior we follow the procedure, suggested in Ref. \cite{CLP}: 
 \beq \label{QSB}
 Q^2_s\Lb b , Y\Rb \,\propto\, \Lb S\Lb b, m \Rb\Rb^{\frac{1}{\bar \gamma}}
 \eeq
 where     $S\Lb b \Rb $ is the Fourier  image of $ S\Lb Q_T\Rb = 1/\Lb 1 +    \frac{Q^2_T}{m^2}\Rb^2$        and the value of $\bar \gamma$ we will discuss below. Such $b$  dependance    results in the large  $b$-dependence of the scattering amplitude,  in the vicinity of the saturation scale which is   proportional to $\exp\Lb - m b\Rb$ at $b \gg 1/m$,  in  accordance of the Froissart theorem \cite{FROI}. In addition, we reproduce the large $Q_T$ dependence of this amplitude proportional to $Q^{-4}_T$ which follows from the perturbative QCD calculation \cite{BRLE}.                                                                                                                                                                                                                                                                                                                                                                                                                                                                                                                                        
                                                                                                                                                                                                                                                                                                                                                                                                                                                                                                                                                              
In building our model we follow the strategy, suggested in Ref. \cite{ IIM}, which  consists  of matching the behavior of the scattering dipole amplitude deep in the saturation domain, that is found using the method  of Ref. \cite{LETU},  and the behavior of the scattering amplitude in the vicinity of the saturation scale \cite{KOLEB,IIML,MUT}.   In this paper, we follow the procedure of Ref. \cite{CLP,CLM} which  allows us to combine  the exact form of the solution inside the saturation domain and in the vicinity of the saturation scale. In   Refs. \cite{SATMOD0,SATMOD1,BKL,SATMOD2,IIM,SATMOD3,SATMOD4,SATMOD5,SATMOD6,SATMOD7,SATMOD8,SATMOD9,SATMOD10, SATMOD11,SATMOD12,SATMOD13,SATMOD14,SATMOD15,SATMOD16,SATMOD17} only the characteristic  behavior of the solution but not the exact form for  it, was used.                                                                                                                                                                                                                                                                                                                                                                                                                                                                                                                                                              
  
  We find  the behavior of the amplitude in the vicinity of the saturation scale,  using the NLO  corrections   to the BFKL Pomeron, calculated in Ref. \cite{BFKLNLO} and the re-summation, suggested in Ref. \cite{SALAM}.  Such behavior has been discussed in Refs. \cite{T,KMRS}. In searching  the parameters of the amplitude we use the procedure\footnote{We  note that this procedure is quite different from the one, used in Ref. \cite{T}. It is  worthwhile mentioning that we do not reproduce the result of Ref. \cite{T}  for energy dependance of the saturation scale, but we are in agreement with  the estimates of Ref. \cite{KMRS} if we apply our calculation to their simplified NLO kernel.}
   , suggested in Ref. \cite{KMRS}, for  full NLO kernel \cite{SALAM} as it has been explored in Ref. \cite{T}.


\section{Theoretical input}
\subsection{General formula}
\begin{figure}[ht]
    \centering
  \leavevmode
      \includegraphics[width=10cm]{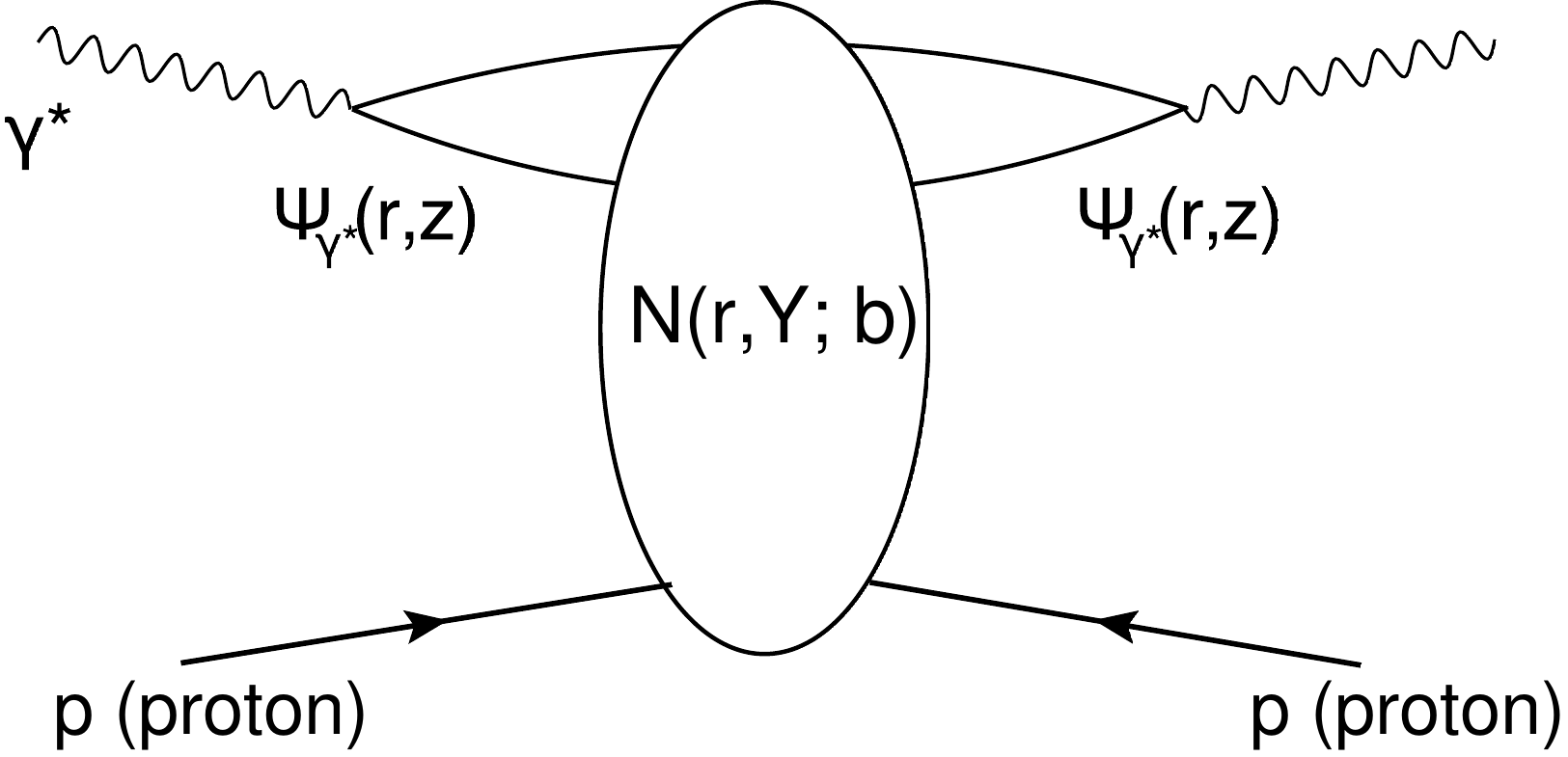}  
      \caption{The graphic representation of \protect\eq{FORMULA} for the scattering amplitude. $Y = \ln\Lb 1/x_{Bj}\Rb$ and $r$ is the size of the interacting dipole.   $z$  denotes  the fraction of energy that is carried by one quark. $b$ denotes  the impact parameter of the scattering amplitude}
\label{gen}
   \end{figure}

The general formula for  deep inelastic processes takes the form (see \fig{gen} and Ref.  \cite{KOLEB} for the review and references therein)

\beq\label{FORMULA}
N\Lb Q, Y; b\Rb \,\,=\,\,\int \frac{d^2 r}{4\,\pi} \int^1_0 d z \,|\Psi_{\gamma^*}\Lb Q, r, z\Rb|^2 \,N\Lb r, Y; b\Rb
\eeq
where $Y \,=\,\ln\Lb 1/x_{Bj}\Rb$ and $x_{Bj}$ is the Bjorken $x$. $z$ is the fraction of energy carried by quark.
$Q$ is the photon virtuality. $b$ denotes  the impact parameter of the scattering amplitude.

\eq{FORMULA} splits the calculation of the scattering amplitude into two stages: calculation of the wave functions, and estimates of the dipole scattering amplitude.

\subsection{Saturation momentum in the NLO}
It is well known that the energy dependance of the saturation momentum can be found from the solution of the linear  BFKL equation \cite{GLR,BALE,IIML,MUT,PEMU}. In the leading order BFKL the saturation momentum $Q_s$ at large values of rapidity has the following form
\beq \label{SM1}
Q^2_s \,\,\propto\,\,e^{\lambda Y}~~~~~\mbox{where}~~~ \lambda \,\,=\,\,\bas \frac{\chi\Lb \gamma_{cr}\Rb}{ 1 - \gamma_{cr}}
~~~~\mbox{and}~~~\chi^{LO}\Lb \gamma \Rb\,=\,2\,\psi\Lb 1 \Rb \,-\, \psi\Lb \gamma\Rb \,-\, \psi\Lb 1 -  \gamma\Rb
\eeq
where $\psi\Lb z \Rb$ is the digamma function.  $\gamma_{cr}$  is the solution of the equation
\beq \label{SM2}
\frac{\chi\Lb \gamma_{cr}\Rb}{1 -   \gamma_{cr}}\,\,=\,\Big{|} \frac{d \chi\Lb \gamma_{cr}\Rb}{d \gamma_{cr}}\Big{|}
\eeq
In the NLO,  the spectrum of the BFKL equation has been found in Ref. \cite{BFKLNLO} and it has the following form:
\beq \label{SM3}
\omega\Lb \gamma\Rb\,\,=\,\,\bas\,\chi^{LO}\Lb \gamma \Rb\,\,+\,\,\bas^2\,\chi^{NLO}\Lb  \gamma\Rb
\eeq
The explicit form of $\chi^{NLO}\Lb  \gamma\Rb$ is given in Ref. \cite{BFKLNLO} (see Appendix 1). However, $\chi^{NLO}\Lb  \gamma\Rb$ turns out to be singular at $\gamma \to 1$,  $\chi^{NLO}\Lb\gamma\Rb \,\propto\,1/(1 - \gamma)^3$. Such singularities indicate that we have to calculate higher order corrections to obtain a reliable result.  The procedure to re-sum high order corrections  is suggested in Ref. \cite{SALAM}. The resulting spectrum of the BFKL equation in the NLO,  can be found from the solution of the following equation \cite{SALAM,T}
\beq \label{SM4}
\omega\,=\,\bas \Lb \chi_0\Lb\omega, \gamma\Rb\,+\,\omega \,\frac{\chi_1\Lb \omega, \gamma\Rb}{ \chi_0\Lb\omega, \gamma\Rb}\Rb
\eeq
where
\beq \label{CHI0}
\chi_0\Lb\omega, \gamma\Rb\,\,=\,\,\chi^{LO}\Lb \gamma\Rb \,-\,\frac{1}{ 1 \,-\,\gamma}\,+\,\frac{1}{1\,-\,\gamma\,+\,\omega}
\eeq
and
\bea \label{CHI1}
&&\chi_1\Lb\omega, \gamma\Rb\,\,=\\
&&\,\,\chi^{NLO}\Lb \gamma\Rb\,+\,F\Lb \frac{1}{1 - \gamma}\,-\,\frac{1}{1\,-\,\gamma\,+\,\omega}\Rb\,+\,\frac{A_T\Lb \omega\Rb \,-\,A_T\Lb 0 \Rb}{\gamma^2} \,+\, \frac{A_T\Lb \omega\Rb - b}{\Lb 1\,-\,\gamma\,+\,\omega\Rb^2}\,-\,\frac{A_T\Lb 0\Rb - b}{\Lb 1\,-\,\gamma\Rb^2}\nn
 \eea
Functions $\chi^{NLO}\Lb \gamma\Rb$ and $A_T\Lb \omega\Rb$ as well as  the constants ($F$ and $b$)  are presented in the Appendix A.

Denoting the solution of \eq{SM4}  $ \omega^{NLO}\Lb \gamma\Rb$  we see that \eq{SM2} for  $\gamma_{cr}$ takes the form
\beq \label{SM5}
\frac{\omega^{NLO}\Lb \gamma_{cr}\Rb}{1 -   \gamma_{cr}}\,\,=\,\Big{|} \frac{d \omega^{NLO}\Lb \gamma_{cr}\Rb}{d \gamma_{cr}}\Big{|}
\eeq
This equation was firstly derived in Ref. \cite{GLR} in the semi-classical approximation for the dipole scattering amplitude. In this approximation the amplitude appears as the wave packet and \eq{SM5} is the condition that the phase velocity of this wave packet  is equal to the group velocity. This condition determines the special line (critical line) which gives the saturation scale. In Refs.  \cite{BALE,KMRS} \eq{SM5} was derived beyond of the semi-classical approximation.

\begin{figure}[h]
    \centering
  \leavevmode
      \includegraphics[width=12cm]{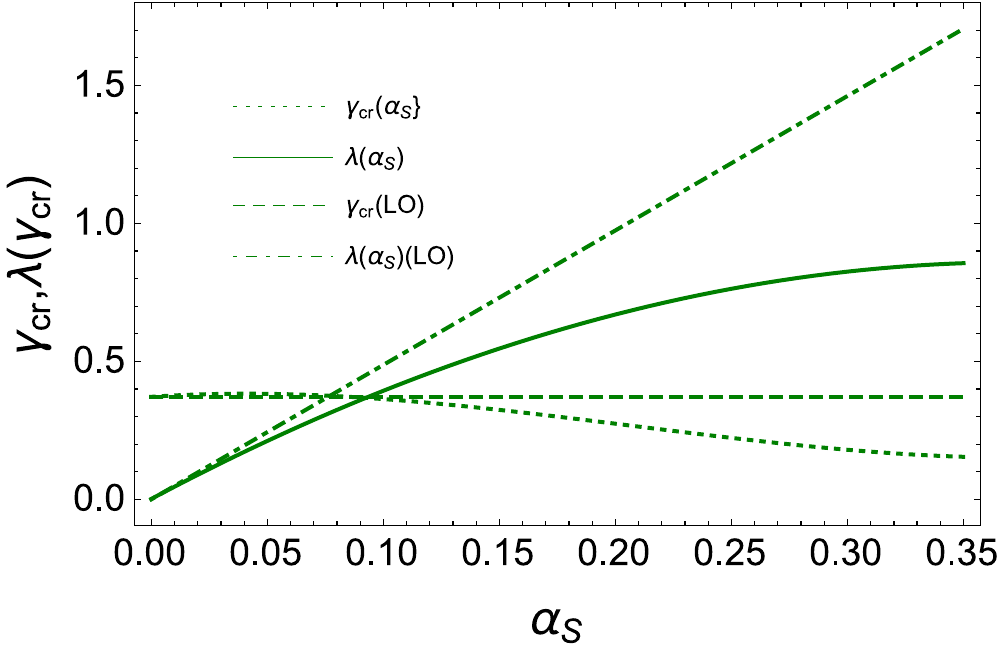}  
      \caption{$\lambda(\gamma_{cr})$ and $\gamma_{cr}$ versus $\as$.}
\label{laga}
   \end{figure}

To find $\omega^{NLO}\Lb \gamma_{cr}\Rb$ and $\gamma_{cr}$ we do not need to solve  \eq{SM3}explicitly . We  can solve the system of two equations:
\bea \label{SM6}
\omega_{cr}\,&=&\,\bas \Lb \chi_0\Lb\omega_{cr}, \gamma_{cr}\Rb\,+\,\omega \,\frac{\chi_1\Lb \omega_{cr}, \gamma_{cr}\Rb}{ \chi_0\Lb\omega_{cr}, \gamma_{cr}\Rb}\Rb\\
\frac{\omega_{cr}}{1 - \gamma_{cr}}\,&=&\,\bas \Lb \chi_0\Lb\omega_{cr}, \gamma_{cr}\Rb\,+\,\omega \,\frac{\chi_1\Lb \omega_{cr}, \gamma_{cr}\Rb}{ \chi_0\Lb\omega_{cr}, \gamma_{cr}\Rb}\Rb'_{\gamma}\Bigg{/}\Bigg(1\,-\,\bas\Lb \chi_0\Lb\omega_{cr}, \gamma_{cr}\Rb\,+\,\omega \,\frac{\chi_1\Lb \omega_{cr}, \gamma_{cr}\Rb}{ \chi_0\Lb\omega_{cr}, \gamma_{cr}\Rb}\Rb'_{\omega}\Bigg)\nn
\eea 
where $\omega_{cr} \equiv \omega^{NLO}\Lb \gamma_{cr}\Rb$.  In \fig{laga} we plot the solution to this set of equations. One can see that both  $\gamma_{cr}$ and  $\lambda$  differ from the leading order estimates.

\begin{figure}[h]
    \centering
  \leavevmode
      \includegraphics[width=8cm]{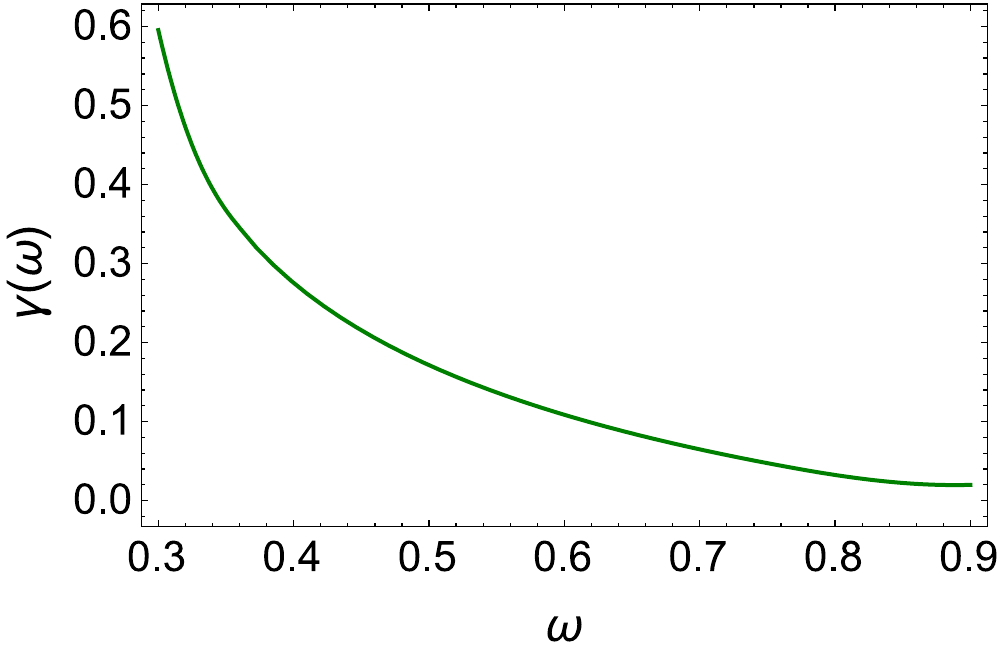}  
      \caption{$\gamma$ versus $\omega  $ for $\bas = 0.15$.}
\label{gavsom}
   \end{figure}

 \fig{gavsom}  shows the solution of \eq{SM4} in the form    $\gamma = \gamma\Lb \omega\Rb$. One can see that $\gamma = \gamma\Lb \omega\Rb \to 0$ at $\omega \to 1$. This property means that we have energy conservation in the NLO,  while in the LO  $\gamma \propto \bas \neq 0$ at $\omega \to 0$,  indicating the energy violation of the order of $\bas$.

\begin{figure}[h]
    \centering
  \leavevmode
      \includegraphics[width=10cm]{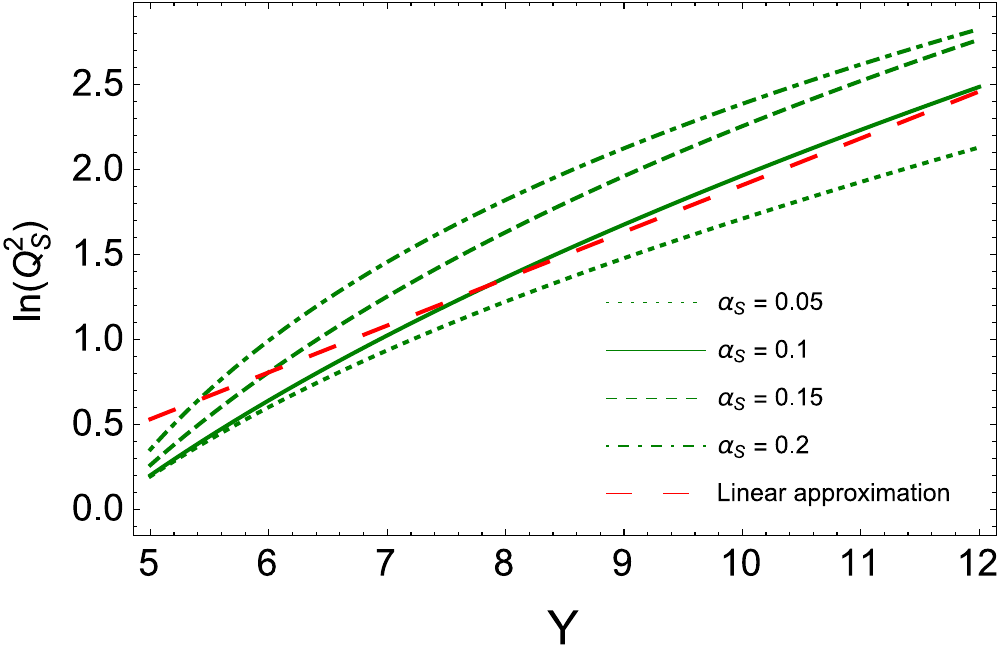}  
      \caption{$\ln\Lb Q_s\Lb b, Y\Rb\Big{/}  Q_s\Lb b, Y=Y_0\Rb\Rb$ versus $Y$ at different values of $\bas$.$ Y_0 = 4.6$. For linear approximation we plot $0.7 \,\lambda_{cr }\Lb Y - Y_0\Rb $ at $\bas = 0.1$.}
\label{lnqs}
   \end{figure}

 The simple energy(rapidity) dependance of \eq{SM1} only holds  at large values of $Y$. The first two corrections lead to  following expression
 
\bea \label{TEQS}
&&\ln\Lb Q^2_s(Y)/Q^2_s(Y_0, b )\Rb\,=\, \lambda^{eff}\Lb \bas, Y, Y_0 \Rb\,\Lb Y \,-\,Y_0\Rb \,\,=\\
&&~~~~~~~~~~~~~ \lambda(\gamma_{cr})\,(Y - Y_0)\,\,
- \,\,\frac{3}{2 ( 1 -
\gamma_{cr})}\,\ln(Y/Y_0)
\, -\,\frac{3}{( 1 - \gamma_{cr})^2}\,\sqrt{\frac{2\,\pi}{\omega''(\gamma_{cr})}}\,
\Lb \frac{1}{\sqrt{Y}}\,-\, 
\frac{1}{\sqrt{Y_0}}\,\Rb\,\,+\,\,{\cal O}\Big( \frac{1}{Y}\Big)\nn
\eea
where $\omega''(\gamma_{cr})= d^2
\omega(\gamma)/(d \gamma)^2$ at $\gamma = \gamma_{cr}$,  the values of $\lambda\Lb \gamma_{cr}\Rb$ and  $\gamma_{cr}$ have been discussed above. $Y_0$ is the value of rapidity from which we start the evolution. The  first term was found in Ref. \cite{GLR}, the second in Ref. \cite{MUT} and the third in Ref. \cite{PEMU}. In \fig{lnqs} $\ln\Lb Q_s\Lb b, Y\Rb\Big{/}  Q_s\Lb b, Y=Y_0\Rb\Rb$ is plotted at different values of $\bas$ in the region of $Y \leq 12$ where the most experimental data are available. In this plot we take into account that  the running QCD coupling has to be taken at scale $Q_s\Lb Y\Rb$ as we will argue in the next section, or in other words we use 
\beq \label{ALHQS}
\bas\Lb Q_s\Rb\,\,\frac{\ba}{1 \,+\,\ba\,b\,\lambda_{cr}\,\Lb Y - Y_0\Rb}
\eeq
where $\ba$ =$\bas\Lb Q_0\Rb $ is the QCD coupling at the scale $Q_0 = Q_s\Lb Y=Y_0\Rb$ (see \eq{QSBP}).

One can  see that the corrections to $\ln\Lb Q_s\Lb b, Y\Rb\Big{/}  Q_s\Lb b, Y=Y_0\Rb\Rb\,\,=\,\,\lambda_{cr}\,\Lb Y - Y_0\Rb$ are essential and they lead to $\lambda^{eff} \,\approx\, 0.7\, \lambda_{cr}$. However, they turn out to be smaller than it was estimated in Ref.\cite{T}, perhaps because the last term in \eq{TEQS} was not taken into account.

\eq{TEQS} shows that while we know the energy dependance theoretically,  the value of $Q^2_s(Y_0, b )$ is our phenomenological input which we will discuss below.


\subsection{Scattering amplitude in the vicinity of the saturation scale}
In the region where $r^2 \,Q^2_s\Lb Y,b\Rb \,\approx\,\,1$ (in the vicinity of the saturation scale)  the scattering amplitude has a well known  behavior  \cite{IIML,MUT,KOLEB}
\beq \label{N1}
N\Lb r, Y; b\Rb\,\,=\,\,N_0 \Lb r^2 \,Q^2_s\Lb b \Rb \Rb^{1 - \gamma_{cr}}
\eeq
where  $\gamma_{cr}$  is the solution to \eq{SM6}.

 The amplitude of \eq{N1} shows a  geometric scaling behavior as a  function of one variable $\tau =  r^2 \,Q^2_s\Lb b \Rb$. Such behavior is proved inside the saturation region  \cite{BALE, LETU} where $ \tau \,\geq\,1$. However, it  actually holds outside of the saturation region for $\tau\,\leq\,1$ \cite{IIML}. In Ref.  \cite{IIML} it  is shown that the first corrections due to a violation of the geometric scaling behavior,  can be taken into account by replacing $1 - \gamma_{cr}$ in \eq{N1}
by the following expression
\beq \label{N2}
1 \,-\,\gamma_{cr}\,\,\to\,\,1 \,-\,\gamma_{cr}\,\,-\,\,\frac{1}{2\,\kappa\,\lambda\,Y} \,\ln\ \Lb r^2 \,Q^2_s\Lb b \Rb \Rb
\eeq
where $\lambda = \bas  \chi\Lb \gamma_{cr}\Rb/\Lb 1 - \gamma_{cr}\Rb$ and $\kappa \,=\,\chi''\Lb \gamma_{cr}\Rb/\chi'\Lb \gamma_{cr}\Rb$.

\begin{boldmath}
\subsection{ The scattering amplitude deep inside  the saturation region ($r^2\,Q^2\Lb b, Y\Rb \,\gg\,1$)}
\end{boldmath}

The non-linear Balitsky-Kovchegov equation has been derived in the NLO, and it takes the form  \cite{NLOBK1,NLOBK2,NLOBK2}

\begin{align}
 \label{BKNLO}
 \hspace*{0cm}
	\frac{\dif S_{12}}{\dif Y}\!= &\frac{\bas}{2\pi}\!
	\int\!\! \dif^2 x_{3}
	\frac{x_{12}^2}{x_{13}^2 x_{23}^2}
	\Bigg\{ 1 \!+\! \bas b
	\left(\ln x_{12}^2 \mu^2 
 \!-\!\frac{x_{13}^2 \!-\! x_{23}^2}{x_{12}^2}
 \ln \frac{x_{13}^2}{x_{23}^2}\right)
 \nn\\
 & \hspace*{-1cm} 
 + \bas \left(\frac{67}{36} \!-\! \frac{\pi^2}{12} \!-\! 
 \frac{5}{18}\,\frac{N_f}{N_c}
 \!-\! \frac{1}{2}\ln \frac{x_{13}^2}{x_{12}^2} 
 \ln \frac{x_{23}^2}{x_{12}^2}
 \right) \Bigg\}
 \left(S_{13} S_{32} \!-\! S_{12} \right)
\nn\\
 & \hspace*{-1cm}+ \frac{\bas^2}{8\pi^2}
 \int \frac{\dif^2 x_3 \,\dif^2 x_4}{x_{34}^4}
 \Bigg\{-2
 + \frac{x_{13}^2 x_{24}^2 + x_{14}^2  x_{23}^2
 - 4 x_{12}^2 x_{34}^2}{x_{13}^2 x_{24}^2 - x_{14}^2 x_{23}^2}
 \nn\\
 & \hspace*{-0.5cm} 
 \ln \frac{x_{13}^2 x_{24}^2}{x_{14}^2 x_{23}^2}
 +\frac{x_{12}^2 x_{34}^2}{x_{13}^2 x_{24}^2}
 \left(1 + \frac{x_{12}^2 x_{34}^2}{x_{13}^2 x_{24}^2 - x_{14}^2 x_{23}^2} \right)
 \ln \frac{x_{13}^2 x_{24}^2}{x_{14}^2 x_{23}^2}
 \Bigg\}
 \left(S_{13} S_{34} S_{42}- S_{13} S_{32} \right)
\end{align}
in \eq{BKNLO} $x_{ik} \,=\,\vec{x}_i \,-\,\vec{x}_j$, $\mu$ is the renormalization scale for the running QCD coupling and all other constants are defined in Appendix A.
$S_{ij}$ is the S-matrix for   scattering of a dipole of size $x_{i j}$,  with the target.

One can see that in the region where $S_{ij} \to 0$,  all terms except the first one,  which is proportional to $S_{12}$, are small and can be neglected. In other words ,  in the region where $S_{12}\,\gg\, S_{13}\,S_{32 }\,\gg\,S_{13} S_{34} S_{42}$ we can reduce \eq{BKNLO} to the following linear equation  \cite{LETU}
\bea
 \label{BKLT}
 \hspace*{0cm}
	&&\frac{\dif S_{12}}{\dif Y}\!= \\
	&& -\,\,\frac{\bas}{2\pi}\!
	\int\!\! \dif^2 x_{3}
	\frac{x_{12}^2}{x_{13}^2 x_{23}^2}
	\Bigg\{ 1 \!+\! \bas b
	\left(\ln x_{12}^2 \mu^2 
 \!-\!\frac{x_{13}^2 \!-\! x_{23}^2}{x_{12}^2}
 \ln \frac{x_{13}^2}{x_{23}^2}\right)
 + \bas \left(\frac{67}{36} \!-\! \frac{\pi^2}{12} \!-\! 
 \frac{5}{18}\,\frac{N_f}{N_c}
 \!-\! \frac{1}{2}\ln \frac{x_{13}^2}{x_{12}^2} 
 \ln \frac{x_{23}^2}{x_{12}^2}
 \right) \Bigg\}
S_{12} \nn
\eea
where $S_{12} \,\equiv\,1\,-\,N\Lb x_{12}, b,Y\Rb$.

The integral over $x_3$ is taken in the Appendix B and \eq{BKLT} can be written in the form
\beq \label{BKLT1}
\frac{\dif \ln S_{12}}{\dif Y}\!=\displaystyle{-\bas \left[ 1+\bas\,b  \ln(\mu^{2} x_{12}^{2})+\bas\left( \frac{67}{36}-\frac{\pi^{2}}{12}-5\frac{N_{f}}{N_{c}} \right) \right]\ln(Q_{s}^{2} x_{12}^{2}) +\frac{\bas^{2} b }{2} \ln^{2}Q_{s}^{2} x_{12}^{2} +\frac{\bas\zeta(3) }{16} }\eeq
In \eq{BKLT1} almost all terms are function of $z \,=\,\ln\Lb x^2_{12}\,Q^2_s\Rb$,  except the term $\bas\,b  \ln(\mu^{2} x_{12}^{2})\ln(Q_{s}^{2} x_{12}^{2})$. Introducing the new renormalization point $Q^2_s$ instead of $\mu^2$ the equations  reduces to the following one
\bea\label{BKLT2}
\frac{\dif \ln S_{12}}{\dif Y}\!&=&\\
 & & \displaystyle{-\bas\Lb Q_s\Rb  \left[ 1+  \frac{3}{2}\bas\Lb Q_s\Rb\,b  \ln(Q_s^{2} x_{12}^{2})+\bas(Q_s)\left( \frac{67}{36}-\frac{\pi^{2}}{12}-5\frac{N_{f}}{N_{c}} \right) \right]\ln(Q_{s}^{2} x_{12}^{2})  +\frac{\bas(Q_s)\zeta(3) }{16} }\nn \eea

Replacing $Y$ by $z  \,=\, \ln \Lb Q^2_s x^2_{12}\Rb\,=\,\bas\Lb Q_s\Rb\, \varrho\, (Y -Y_0)$ \eq{BKLT2} takes the form

\beq\label{BKLT3}
\frac{\dif \ln S_{12}}{\dif z}  =
 \displaystyle{-\frac{1}{\varrho} \Bigg( \left[ 1+  \frac{3}{2}\bas\Lb Q_s\Rb b  z +\bas(Q_s)\,\left( \frac{67}{36}-\frac{\pi^{2}}{12}-5\frac{N_{f}}{N_{c}} \right) \right] z   +\frac{\zeta(3) }{16} }\Bigg)\eeq

 Integration over $z$ leads to
 \beq\label{BKLT4}
 \ln S_{12} =
 \displaystyle{-\frac{1}{2\,\varrho} \Bigg( \left[ z +  \bas\Lb Q_s\Rb b  z^2  +\bas(Q_s)\,z\,\left( \frac{67}{36}-\frac{\pi^{2}}{12}-5\frac{N_{f}}{N_{c}} \right) \right] z   +\frac{\zeta(3) }{8} }\,z \Bigg) \eeq

\eq{BKLT4} shows a geometric scaling behavior,  being  a function of one variable 	$z$. However, this scaling behavior only holds,  if we choose the renormalization scale $\mu = Q_s$.

Finally,
 \bea\label{BKLT5}
1 -  N\Lb z\Rb \, &=&\, e^{ - {\cal Z }\Lb z \Rb}  \\
\mbox{with}~ & &{\cal Z}\Lb z \Rb\,=\,
\frac{1}{2 \,\varrho} \,\Bigg( \left[ z +  \bas\Lb Q_s\Rb b  z^2  +\bas(Q_s)\,z\,\left( \frac{67}{36}-\frac{\pi^{2}}{12}-5\frac{N_{f}}{N_c}\right) \right] z -
\,\frac{\zeta(3)}{8} \, z\Bigg) \nn
   \eea
   where we replace $\varrho$ by $ \varrho \,=\,\lambda_{cr}\Lb \bas \Rb/\bas$.

\begin{boldmath}
\subsection{ Matching at $r^2\,Q^2\Lb b, Y\Rb\, = \,1$}
\end{boldmath}


In section 2.3 we  saw that the amplitude in the vicinity of the saturation scale  has a geometric scaling behavior ( see \eq{N1})  as well as the amplitude at $r^2 Q^2_s \gg 1$,  as has been shown in the previous section. The first observation is that we can match these two amplitude, only if we assume that the renormalization scale $\mu = Q_s$. Practically , it means that  we have to replace $\bas$ in section  2.3  by $\bas\Lb Q_s\Rb$ . This  generates an additional $Y$ dependence, diminishing the value of $\lambda_{cr}$ at large values of $Y$. 

The general matching conditions have the form of   two following  equations at $z = z_m$:
We match these two solution at $z = z_{m}$ where
\beq \label{MC}
 N^{0 < z \ll 1}\Lb z = z_m \Rb \,=\, N^{z \gg  1}\Lb z = z_m \Rb;~~~~~~~~~~~ \frac{d N^{0 <z \ll 1}\Lb z = z_m \Rb}{d z_m} \,=\, \frac{d N^{z \gg  1}\Lb z = z_m \Rb}{ d z_m};~ 
 \eeq

     \begin{figure}[ht]
  \begin{tabular}{ccc}
      \includegraphics[width=5.5cm]{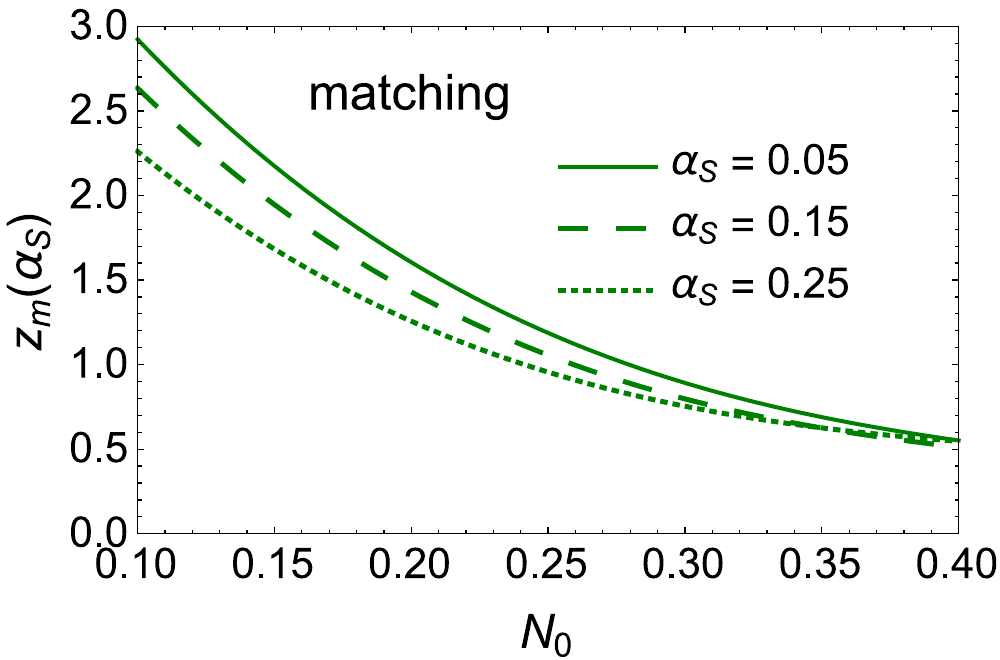} &   \includegraphics[width=5.5cm]{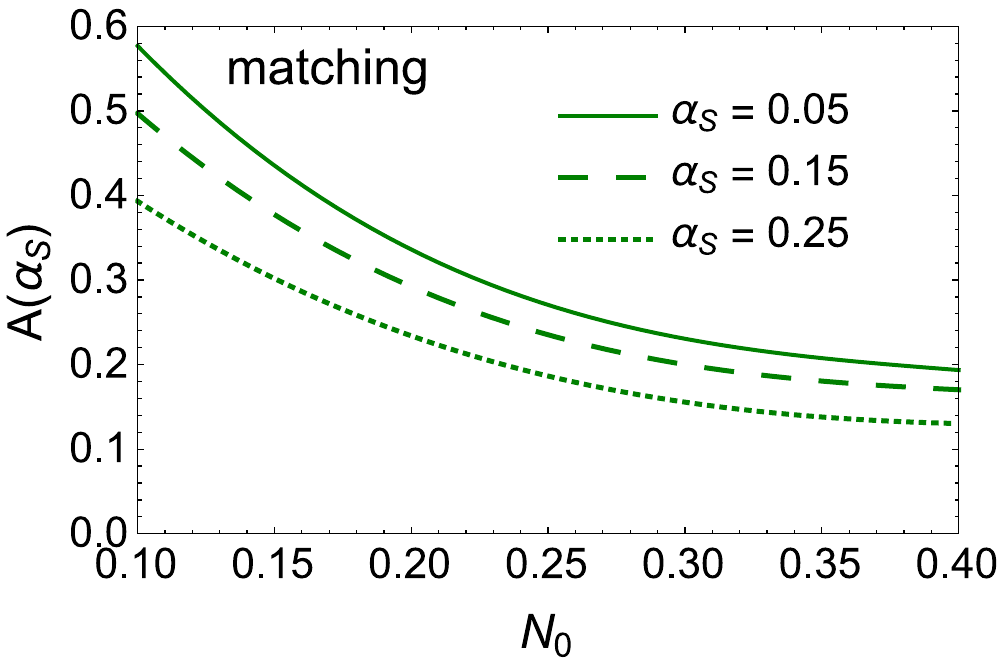} &  \includegraphics[width=5.5cm]{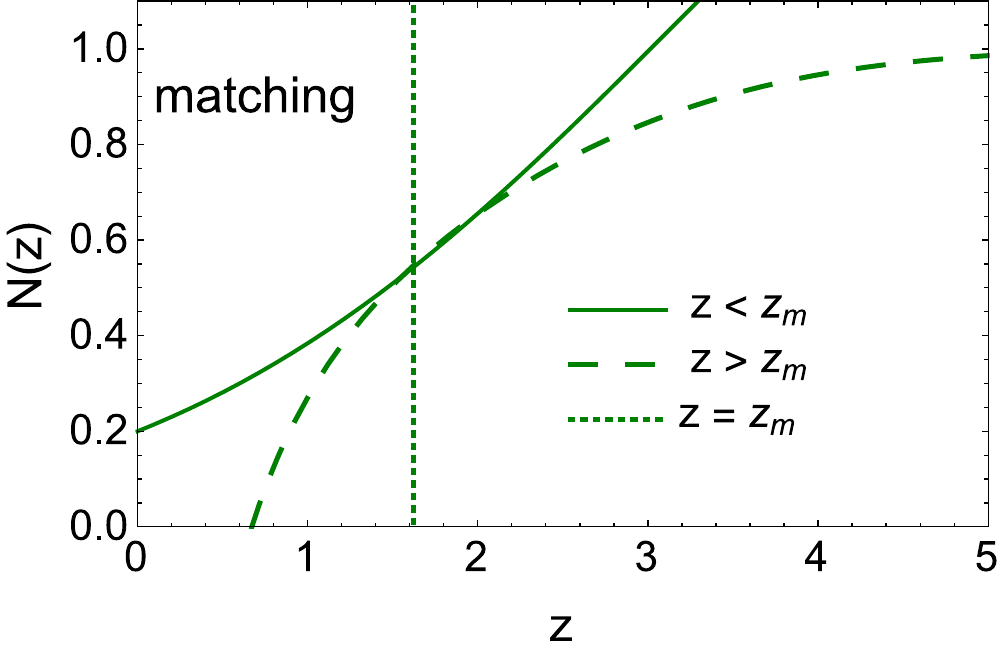}\\
      \fig{match}-a&\fig{match}-b & \fig{match}-c\\
      \end{tabular}
      \caption{Matching procedure:  function $z_m\Lb N_0, \bas\Rb$ (\fig{match} -a), function $A\Lb N_0, \bas\Rb$ (\fig{match}-b) and the example of the resulting function for $N_0=0.1$ and $\bas=0.15$  ( \fig{match}-c) .}
\label{match}
   \end{figure}

These two equations determined the value of the amplitude and the point of matching. The additional restriction is that $z_m \ll 1$ , or,  in other words $z_m$ should be in the vicinity of the saturation scale. A problem is  that it  is impossible to satisfy  \eq{MC}  without modifying  the solution of \eq{BKLT5}. Most models in the past followed the suggestion of Ref. \cite{IIM} and instead of \eq{BKLT5}, the modified solution 
\beq
1 \,- \, N\Lb z\Rb \, =\,e^{ - C {\cal Z }\Lb z \Rb}
\eeq
was introduced,  in  which the value of constant $C$ was  determined by the matching conditions of \eq{MC}. In Ref. \cite{CLM}  the correction to the asymptotic solution of \eq{BKLT5} was found,  which allows us to use the solution of \eq{BKLT5} without an arbitrary unjustified constant $C$. This  solution takes the form
\bea \label{MF}
N^{z \,\gg\,1} \Lb z \Rb &\,=\,&1 - 2\,A e^{- {\cal Z}}\,\,- \,A^2\,\frac{1}{{\cal Z} }\,e^{- 2 {\cal Z}}\,\,+\,\,{\cal O}\Lb e^{-3 {\cal Z}}\Rb\\
{\cal Z}&=&{\cal Z}\Lb \eq{BKLT5}, z \to  z -  \h A \sqrt{
\varrho\, \pi/2} - 2\psi(1)\Rb\nn
\eea
where $\psi(x)$ is the digamma function (see Ref. \cite{RY} formula {\bf 8.360 - 8.367}).

The second term in \eq{MF} is the solution given in  Ref. \cite{LETU},  in which the theoretically unknown constant $A$ is introduced,  both as the coefficient in front, and as correction to the argument. The third term is the next order correction  at large $z$. In Ref. \cite{CLM,CLP} it has been demonstrate that using \eq{MF},  we can solve \eq{MC}
and  find $z_m $.
  \subsection{Impact parameter dependance of the saturation scale} 
So far we have introduced only one phenomenalogical parameter $N_0$: the value of the scattering amplitude at $r^2\,Q^2_s = 1$. However, we need to 
specify the value of the saturation scale at $Y = Y_0$.   It includes the value of the saturation scale and its dependence on the impact parameter $b$.
Both 
 can only  be estimated in non-perturbative QCD.  Due to the embryonic stage of our  understanding of non-perturbative QCD contribution,  we can only suggest a phenomenological parameterization.
 
 For  $Q_s\Lb Y = Y_0,b\Rb$ we use the following expression
 \beq \label{QSBP}
   Q^2_s\Lb Y = Y_0,b\Rb\,\,=\,\,Q^2_0\,S\Lb b \Rb\,\,=\,\, Q^2 _0
   \,\Lb m\,b\,K_1\Lb m\, b\Rb\Rb^{1/(1 - \gamma_{cr})}
   \eeq

   The value of $m$ has to be find from the fitting of the experimental data. We expect  that $m \approx 0.5 \div  0.85\,GeV$ since $m = 0.72\,GeV$ is the scale for the electromagnetic form factor of proton, while $m \approx 0.5\,GeV$ is the scale for so called gluon mass \cite{GLMASS}.

 We differ  from other models in that \eq{QSBP} leads to  $   Q^2_s\Lb Y = Y_0,b\Rb\,\xrightarrow{m b \,\gg\,1}
 \exp\Lb - m b/(1 - \gamma_{cr})\Rb$,   providing the correct large $b$ behavior of the scattering amplitude. It should be stressed that the exponential decrease at large $b$,  follows from a general theoretical approach,  based on analyticity and unitarity of the scattering amplitude (see  \cite{FROI}). Therefore,  $Q^2_s\Lb Y = Y_0,b\Rb\,\propto \exp\Lb- b^2/B\Rb$ that was used in other models 
  (see Refs. \cite{SATMOD5,SATMOD6,SATMOD7,SATMOD8,SATMOD9,SATMOD12,SATMOD17})   are in the direct contradiction with theory.
  The behavior of the amplitude at large $b$  determines the energy dependance of the interaction radius, leading to $R \propto (1/m) Y$ for the exponential decrease,  and $R \propto (1/m) \sqrt{Y}$ for the Gaussian $b$ dependance. Such a difference , leads  to a fast increase of the scattering amplitude for our parameterization and it will effect the predictions at high energy.
  
  \eq{QSBP} gives the amplitude in the vicinity of the saturation scale, which is proportional to $S\Lb b \Rb$ and generates the behavior $ 1/\Lb 1 + \frac{Q^2_T}{m^2}\Rb^2$, where $Q_T$ is the momentum transfer.  At  large $ Q_T$ the amplitude in our parameterization is proportional ($ A \propto 1/Q^4_T$)   as it follows from the perturbative QCD calculation  \cite{BRLE},  but cannot be reproduced with the Gaussian distribution.

\subsection{Wave functions}
The wave function in the master equation (see \eq{FORMULA}) is the main source of theoretical uncertainties: even in the case of  deep inelastic processes,  we can trust the wave function of perturbative QCD only, at rather large values of $Q^2 \,\geq\,Q^2_0$ with $Q^2_0 \approx 0.7 GeV^2$ (see Ref.  \cite{GLMTC}). The expression for $(\Psi^*\Psi)^{\gamma^*} \equiv \Psi_{\gamma^*}\Lb Q, r, z\Rb \,\Psi_{\gamma^*}\Lb  Q, r,z\Rb$ is well known (see Ref. \cite{KOLEB} and references therein)
\begin{align}
  (\Psi^*\Psi)_{T}^{\gamma^*} &=
   \frac{2N_c}{\pi}\alpha_{\mathrm{em}}\sum_f e_f^2\left\{\left[z^2+(1-z)^2\right]\epsilon^2 K_1^2(\epsilon r) + m_f^2 K_0^2(\epsilon r)\right\},\label{WFDIST}   
  \\
  (\Psi^*\Psi)_{L}^{\gamma^*}&
  = \frac{8N_c}{\pi}\alpha_{\mathrm{em}} \sum_f e_f^2 Q^2 z^2(1-z)^2 K_0^2(\epsilon r),
\label{WFDISL}
\end{align}
where T(L) denotes the polarization of the photon and $f$ is the flavours of the quarks. $\epsilon^2\,\,=\,\,m^2_f\,\,+\,\,Q^2 z (1 - z)$.

\section{Fitting $F_2$ and values of the parameters}
The most accurate  experimental data  available  are for  the deep inelastic structure function $F_2$  \cite{HERA1}, which we will attempt to fit using the model.
 As has been mentioned, we can trust our model in the restricted kinematic region, which we choose in the following way: $ 0.85\,GeV^2 \leq Q^2\leq 60\,GeV^2$ and  $x \,\leq \,0.01$. The lower limit of $Q^2$ stems from non-perturbative correction to the wave  function of the virtual photon, while the upper limit originates  from the restriction $x\,\leq\,0.01$. This restriction can be translated to the value of $Y_0$ in our theoretical formulae leading to $Y_0 = 4.6$. Actually we view $Y_0$ as the parameter of the fit (see Table 1).


\begin{table}[ht]
{\small
\begin{tabular}{||l|l|l|l|l|l|l|l|l|l||}
\hline
\hline
$\ba $ & $N_0$ & $Y_0$ & m ($GeV$)& $Q^2_0$ ($GeV^2$) & $m_u $(MeV) &  $m_d $(MeV) &  $m_s $(MeV)&  $m_c $(GeV)  &$\chi^2/d.o.f.$ \\
\hline
0.133&  0.1075 & 3.77 & 0.83&  3.0 & 2.3 &4.8& 95&1.4& 183/153 =1.2\\
\hline
0.143 & 0.0915 & 3.73 & 0.67&2.6&  140 &140 &140& 1.4 & 242/153 = 1.58\\
\hline\hline
\end{tabular}}
\caption{Parameters of the model.    $\ba $, $N_0$,  $m$ and  $Q^2_0$ are fitted parameters. The masses of quarks are chosen  as they  are shown in the table. Two sets  are  related  to two choices of the quark masses: the current masses and the masses of light quarks  are equal to  $140 \,MeV$ which is the typical infra-red cutoff in our approach. }
\label{t1}
\end{table}
 Energy dependance of the saturation scale $Q_s$ and  $\tau = r^2\,Q^2_s\Lb b, Y\Rb$ dependance of the scattering amplitude
 are determined by \eq{TEQS} and \eq{MC}. One can see that both depend on $\bas\Lb Q_s\Rb $   for which we use   \eq{ALHQS}. From this equation  one can see  that we have two fitting parameters: $\ba$ and $Q^2_0$.  In principle, $\ba$ is the running QCD coupling at $Q^2=Q^2_0$, but we consider both $\ba$ and $Q^2_0$ as independent fitting parameters, since we do not want to fix the value of $\Lambda_{\rm QCD}$.  We have two dimensional parameters:
 $Q_0$, which determines the value of $Q_s^2$, and $m$ which determines  its dependance on impact parameters $b$ ( see section 2.6). $N_0$ is  the value of the scattering amplitude at $\tau = 1$. In principle,  the value of $N_0$ can be calculated   using the linear evolution equation with  the initial conditions. However, it  depends on the phenomenological parameters of this initial condition. So we choose $N_0$ as a fitting parameter. 
 
\begin{figure}[h]
    \centering
  \leavevmode
      \includegraphics[width=10cm]{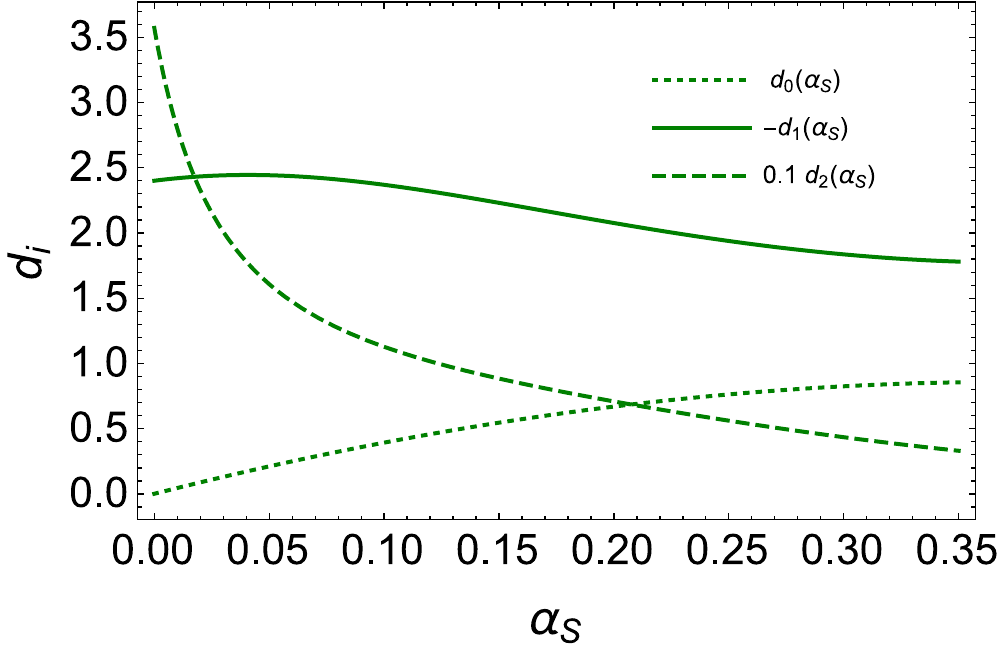}  
      \caption{Function $d_i(\bas)$ of \protect\eq{FIT1} versus $\bas$.}
\label{d}
   \end{figure}

 It is worth mentioning that $\lambda_{cr}, \gamma_{cr}$ are not the fitting parameters as they are in the leading order models. We recall that
 \beq \label{FIT1}
 \ln\Lb Q^2_s\Lb b, Y\Rb\Big{/}  Q^2_s\Lb b, Y=Y_0\Rb \Rb\,=\,d_0\Lb \bas\Rb \, Y \,\,+\,d_1\Lb \bas\Rb \ln\Lb Y/Y_0\Rb\,-\,d_2\Lb \bas
 \Rb \Lb \frac{1}{\sqrt{Y}} \,-\,\frac{1}{\sqrt{Y_0}} \Rb
 \eeq
 where function $d_i$ are shown in \fig{d}.
 In \eq{FIT1} $Y = \ln\Lb 1/x\Rb$,  where $x $ is the Bjorken $x = Q^2/s$ for the deep inelastic scattering with the light quarks ( $Q$ is the photon virtuality and $s$ is the energy squared of collision).  For the charm quark we consider $Y_c = \ln\Lb 1/x_c\Rb$ with $x_c\,=\,(1 + 4 \,m^2_c/Q^2)\,x$.
 
  We do not regard  masses of the quark   as fitting parameters and consider two sets of these masses. In the first set we take the current masses (see the first row of Table 1), and we consider this as the most reliable fit,  based on the consistent theoretical approach. It should be mentioned that for the description of the interaction with $c$-quark we use $Y = \ln \Lb 1/x_c\Rb$ with $x_c = x/(1 + 4\,m^2_c/Q^2)$.
   We  also make  a fit  putting all masses of light quarks (second row of Table 1) to be equal to 140 MeV. We view this mass as a typical infra-red cutoff that we  introduce to take into account the unknown mechanism of confinement.
~

\begin{figure}[h]
    \centering
  \leavevmode
      \includegraphics[width=14cm]{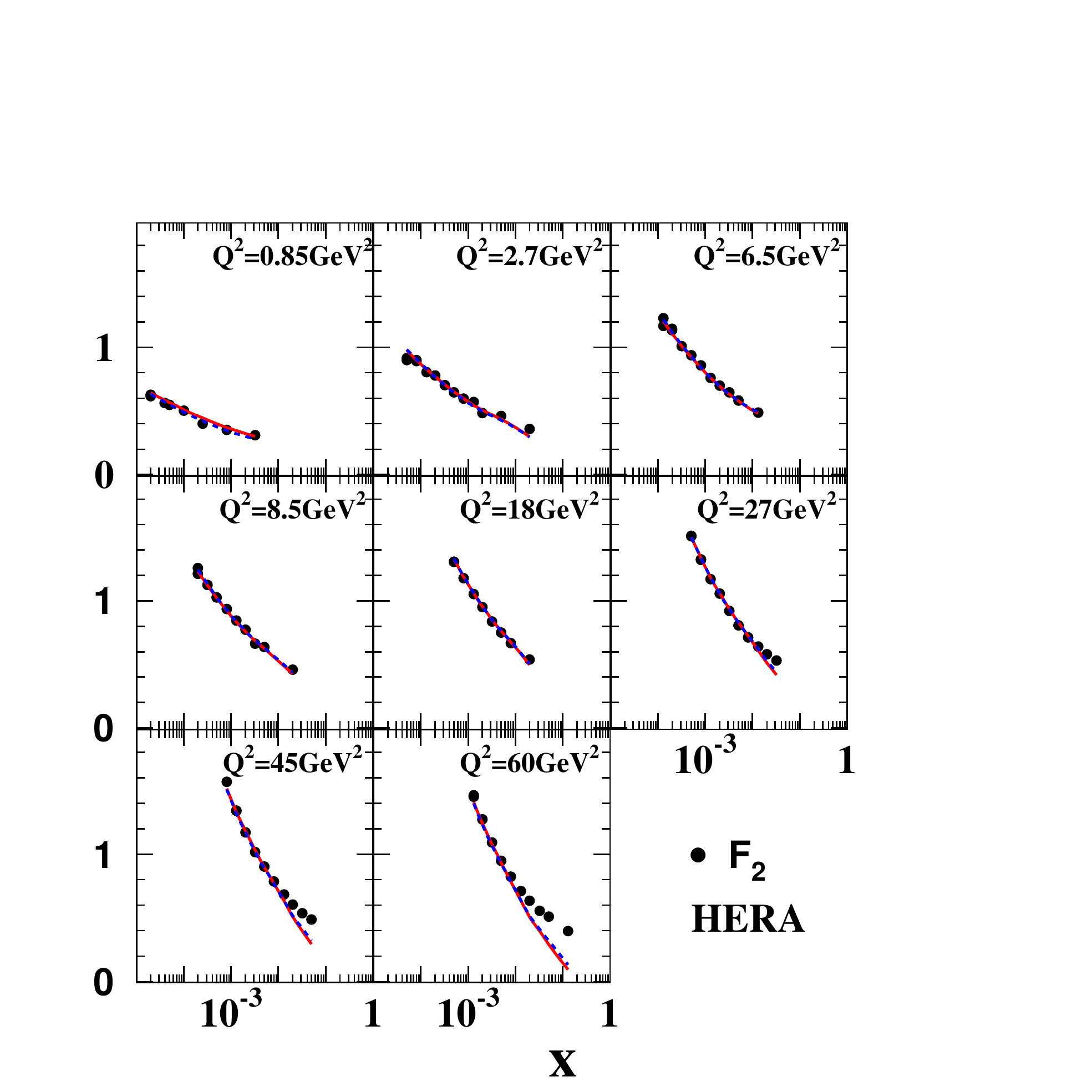}  
      \caption{Our fit of $F_2$ with the values of  parameters given by Table 1. The first set of parameters is shown in solid red curves while the second in blue dotted  lines. The data is taken from Ref. \cite{HERA1}.}
\label{fit}
   \end{figure}

\begin{figure}[h]
    \centering
  \leavevmode
      \includegraphics[width=10cm]{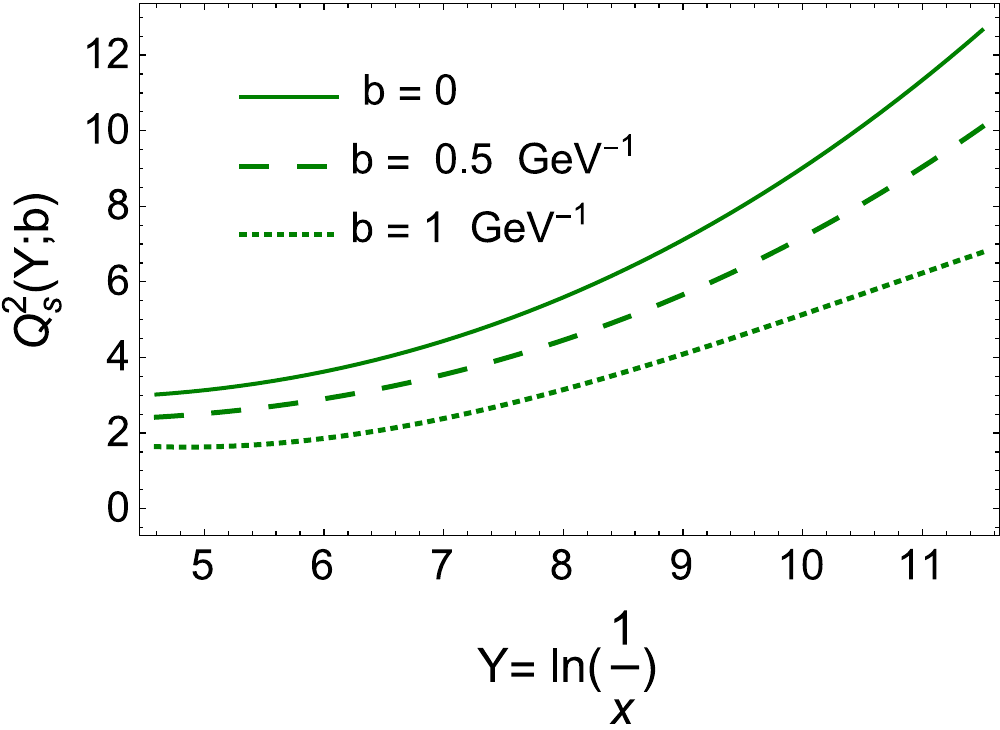}  
      \caption{The value of the saturation momentum $Q^2_s\Lb x, b\Rb$ versus $x$ at fixed $b$ for the  parameters given by Table 1. }
\label{qs}
   \end{figure}

Table 1 gives the values of the fitting parameters, and \fig{fit} demonstrates the quality of the fit. One can see that we describe the data quite well but we have to admit that the quality of the fit is worse than in our model based on leading order QCD estimates\cite{CLP}, in which we fitted the value of $\lambda_{cr}$. $\chi^2/d.o.f = 1.3$ in this fit against $\chi^2/d.o.f = 1.15$ in the fit of Ref.\cite{CLP}.  However, the main complication of this model is that it gives rather a  large value of $Q_0^2$ (see \fig{qs}) which is in sharp contradiction to  the value of the saturation momentum, from all other model description of the experimental data\cite{CLP,SATMOD0,SATMOD1,BKL,SATMOD2,IIM,SATMOD3,SATMOD4,SATMOD5,SATMOD6,SATMOD7,SATMOD8,SATMOD9,SATMOD10, SATMOD11,SATMOD12,SATMOD13,SATMOD14,SATMOD15,SATMOD16,SATMOD17}. The large value of $Q^2_0$ is in agreement with small values of $\ba$,  we  note that $\ba = 0.28$ for $\Lambda_{\rm QCD} = 158 \,MeV$ instead of $\ba = 0.13$ from our fit.

\begin{figure}[h]
    \centering
  \leavevmode
      \includegraphics[width=12cm]{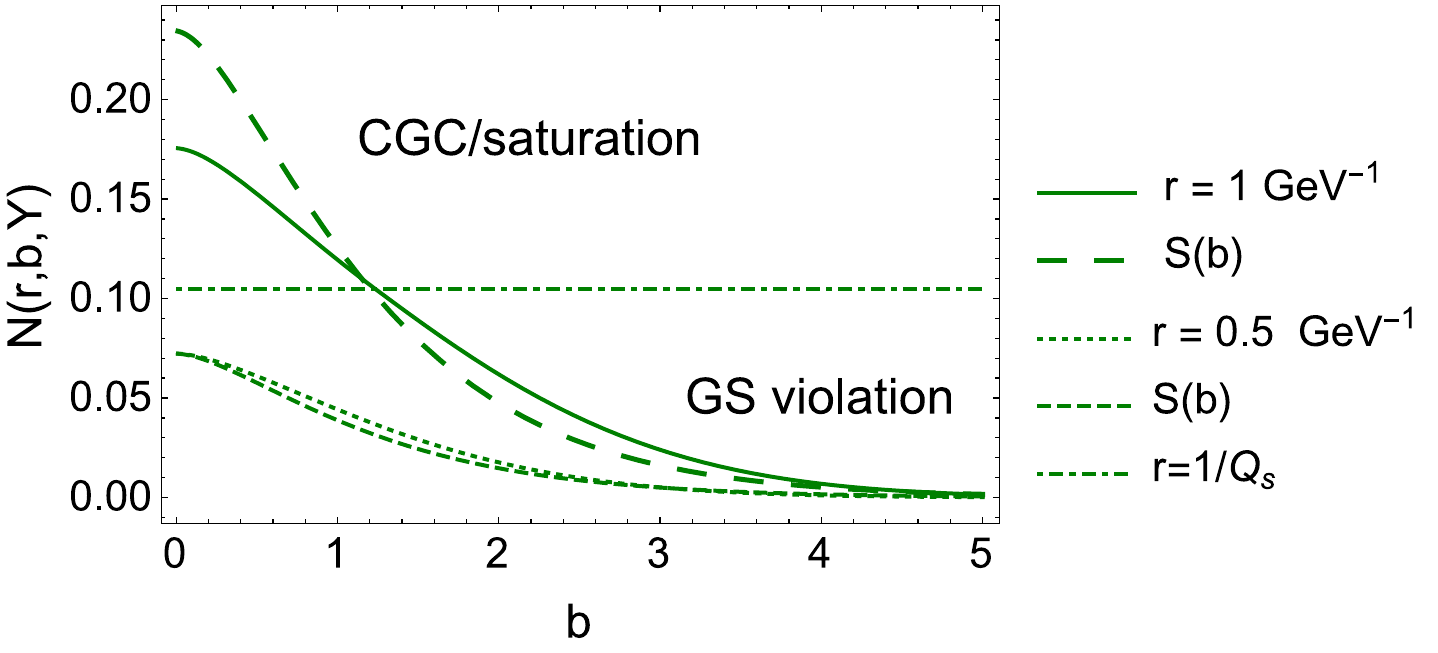}  
      \caption{The  $b$-dependence of the scattering amplitude for the  parameters given by Table 1. $S\Lb b\Rb$ is given by \protect\eq{QSB}.}
\label{nb}
   \end{figure}

The value of $m$ is larger than the typical mass in the electro-magnetic form factor of the proton, but we do not expect that it to  be the same.  Note that the decrease of $Q^2_s$ at large $b$ is proportional to $\exp\Lb - \frac{m}{1 - \gamma_{cr}}\,  b\Rb = \exp\Lb - 1.6 \Lb GeV^{-1}\Rb\,b\Rb$. On the other hand the  behavior of amplitude  on $b$ differs from 
 the saturation scale. In \fig{nb} one can see that both the saturation,  and the violation of the geometric scaling behavior influence the resulting b-dependence of scattering amplitude. Saturation flattens the $ b$-dependence at small values of $b$, while the large $b$ behaviour shows a more rapid decrease than the $b$-dependence of the saturation scale  (see \fig{nb}).

%
It should be stressed that in the framework of our parametrization of the $b$-dependence of the saturation momentum, the scattering amplitude decreases as $\exp\Lb - m b\Rb$ while in all other models on the market it has a Gaussian behavior: $\exp\Lb - m^2\,b^2\Rb$.
     \begin{figure}[h]
    \centering
  \leavevmode
      \includegraphics[width=14cm]{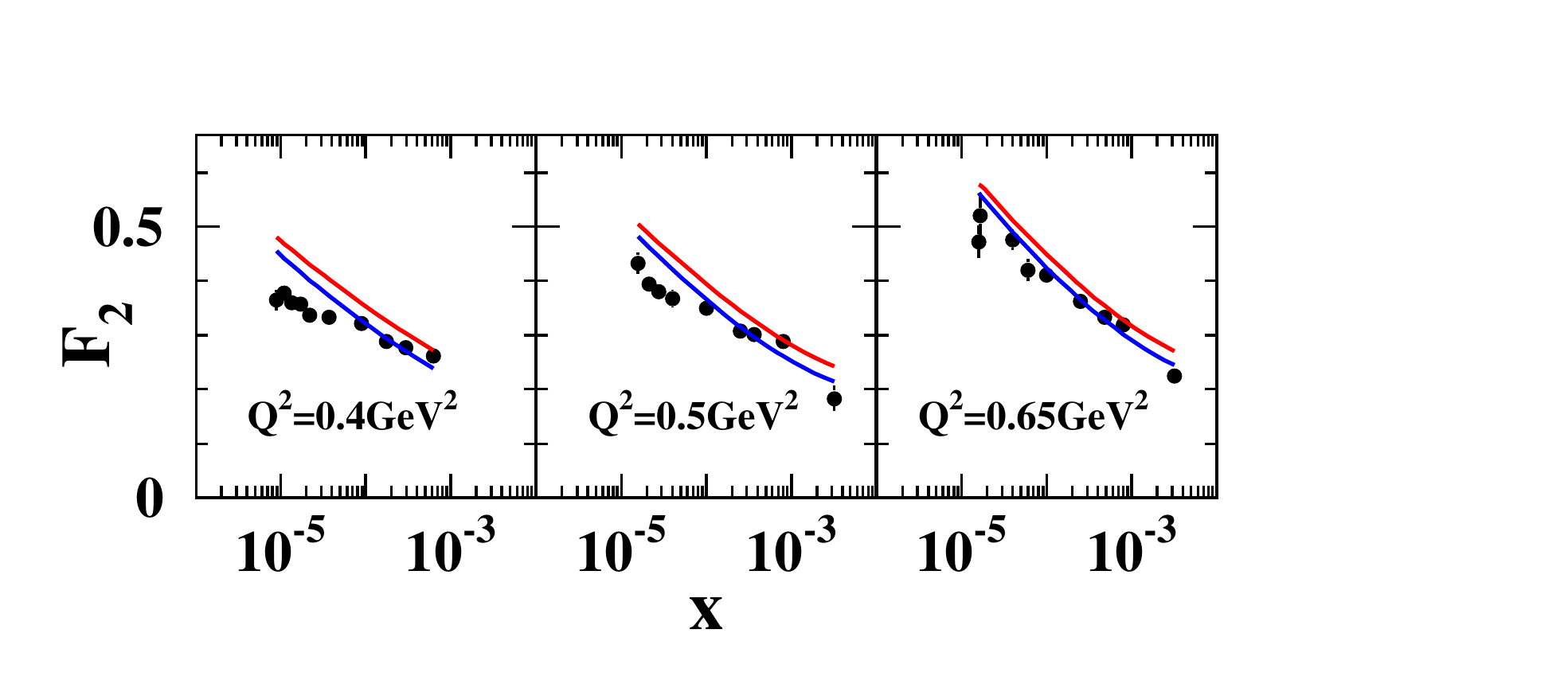}  
      \caption{The  $x$-dependence of $F^{c\bar{c}}_2$ at small  values of $Q^2 \,<\,0.85\,GeV^2$ for the  parameters given by Table 1. The red (upper)  line corresponds to set 1(upper row  of  Table 1)  while  the blue one (low)  is the description with set 2. The data are taken from Ref.\protect \cite{HERAFL1,HERAFL2}.}
\label{f2sq}
   \end{figure}

 
   \fig{f2sq} we present the comparison between our fit of $F_2$ with two sets of parameters at low values of $Q$.
The set with large masses of quarks leads to a much better description illustrating the the non-perturbative corrections to the wave function of the virtual photon are essential at $Q^2 \,<\,0.85 \,GeV^2$.

\begin{figure}[h]
    \centering
  \leavevmode
      \includegraphics[width=14cm]{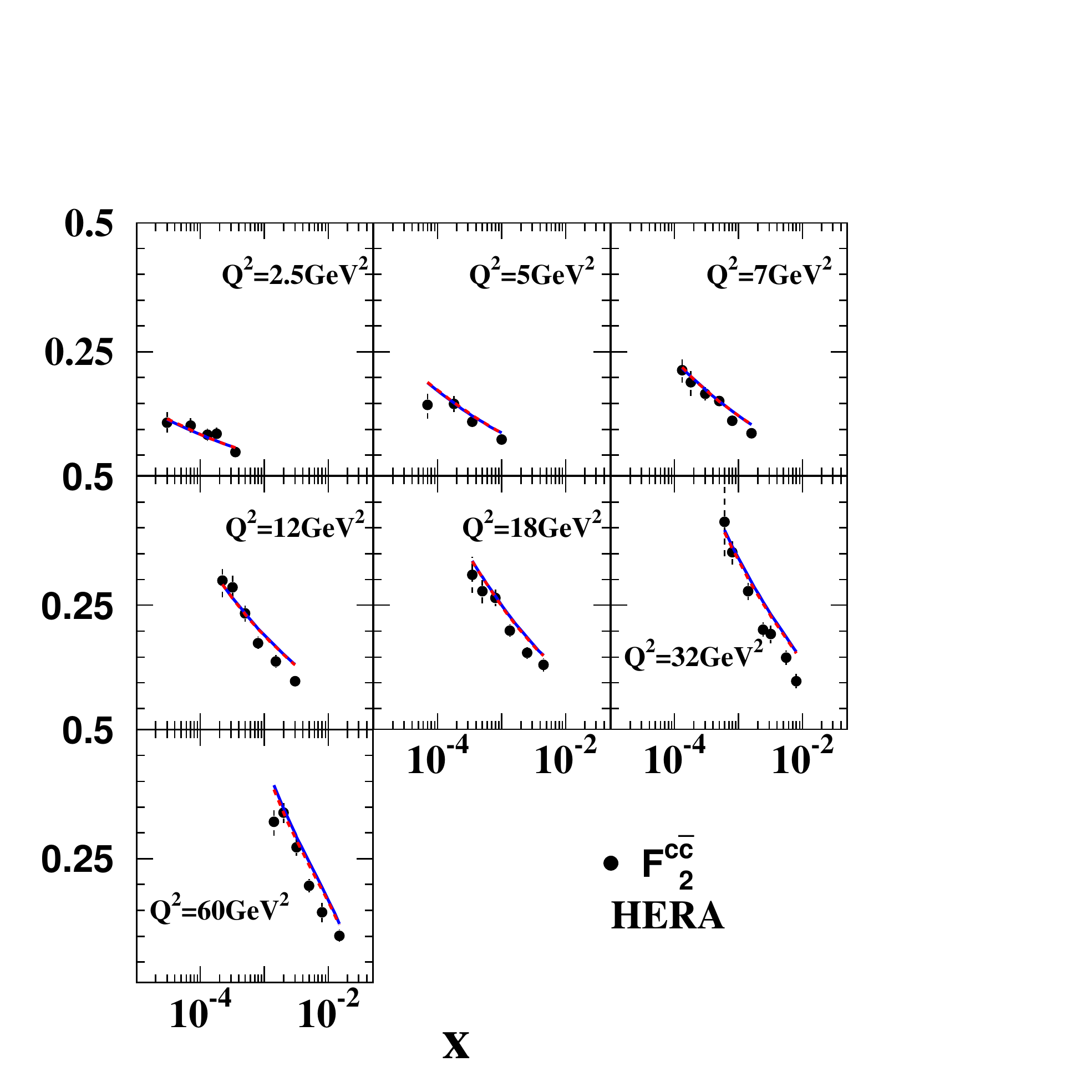}  
      \caption{The  $x$-dependence of $F^{c\bar{c}}_2$ at fixed values of $Q^2$: $ 0.85\,\leq\, Q^2\,\leq\,60\, GeV^2$ for the  parameters given in Table 1. The data are taken from Ref. \protect \cite{HERA2}.}
\label{cc}
   \end{figure}

 $ \mathbf{F_2^{c\bar{c}} } $ : The contribution of the $ c \bar{c}$ pair to the deep inelastic structure function can be calculated with the same theoretical accuracy as the inclusive $F_2$. In \fig{cc} we compare the HERA data on $F^{cc}_2$   \cite{HERA2} with the theoretical predictions.
 One can see that the agreement is reasonable.
   
 ~

\begin{figure}[h]
    \centering
  \leavevmode
      \includegraphics[width=14cm]{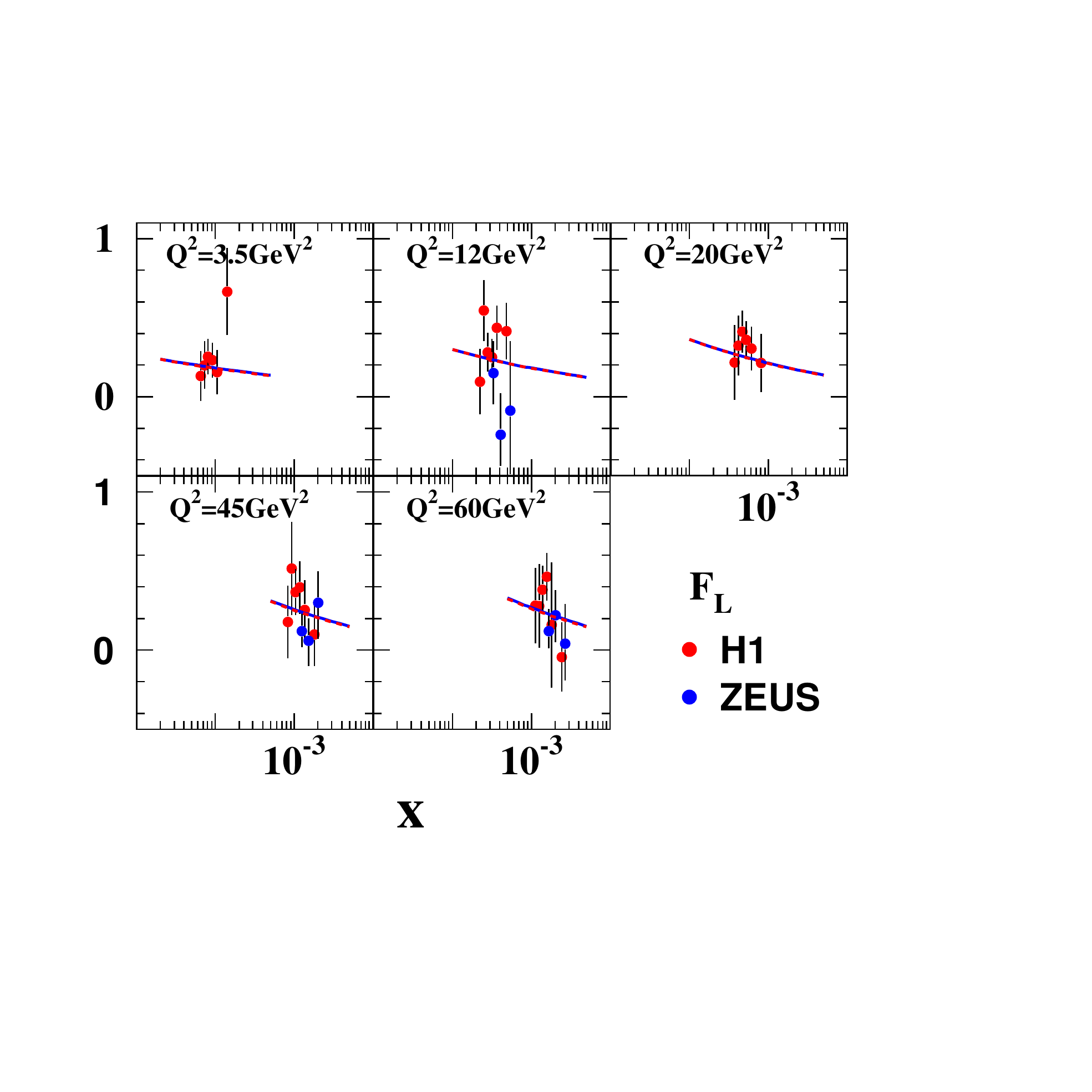}  
      \caption{The  $x$-dependence of $F_L$ at fixed values of $Q^2$: $ 0.85\,\leq\, Q^2\,\leq\,60\, GeV^2$ for the  parameters given in Table 1. The red (blue) lines correspond to set 1 and set 2 fits. The data are taken from Ref. \protect \cite{HERAFL2}.}
\label{f2l}
   \end{figure}

   $ \mathbf{F_L } $ :  $F_L$ can be calculated to the same accuracy as  $F^{c\bar{c}}_2$,  and the comparison with the scant data  available  \cite{HERAFL1,HERAFL2} is plotted in \fig{f2l}. Two sets produce  the same quality of the descriptions since the values of $Q$ are rather large.

        \section{Conclusions}
       

 In this paper we make the first attempt to include everything,  that we have learned about the next-to-leading corrections of  perturbative QCD, into the CGC/saturation model.  In the paper we obtained  two new theoretical  results: (i)  using the approach suggested in Ref.\cite{LETU}, we obtained asymptotic behaviour of the solution to the Balitsky - Kovchegov equation in the NLO of perturbative QCD  \cite{NLOBK1,NLOBK2,NLOBK2}  deep inside of the saturation domain; and (ii) the geometric scaling behaviour of the scattering amplitude,  which holds only if  $\tau = r^2\,Q_s^2\Lb Y; b\Rb$ is  determined in pQCD with the renormalization scale    $Q_s\Lb Y; b\Rb$.  
  
 In the model we include   several known  ingredients: (i) the behaviour of the scattering amplitude in the vicinity of the saturation momentum, using the NLO  BFKL kernel; (ii) the pre-asymptotic behaviour of $\ln\Lb Q^2_s\Lb Y \Rb\Rb$ as function of $Y$ and (iii)  the impact parameter behaviour of the saturation momentum  which has exponential behaviour $\propto \exp\Lb - \,m\, b\Rb$ at large $b$.
 
 In comparison with the models on the market \cite{SATMOD0,SATMOD1,BKL,SATMOD2,IIM,SATMOD3,SATMOD4,SATMOD5,SATMOD6,SATMOD7,SATMOD8,SATMOD9,SATMOD10, SATMOD11,SATMOD12,SATMOD13,SATMOD14,SATMOD15,SATMOD16,SATMOD17},  we add the NLO corrections both deep in the saturation domain and in the vicinity of the saturation scale, as well as  
     two crucial ingredients follow Ref.\cite{CLP}:  the correct solution to the non-linear (BK) equation  \cite{BK} in the saturation region, and   impact parameter distribution that leads to
 exponential decrease of the saturation momentum at large impact parameters  and to  power-like decrease at large transfer momentum that follows from perturbative  QCD.\cite{BRLE}.
 
 In spite of the fact that we describe the experimental data fairly we are aware that our description is worse than in the CGC/saturation models based on the leading order QCD approach.  The main difficulties are related to the small value of the QCD coupling at $Q_s\Lb Y_0\Rb$, and the large values of the saturation momentum, which show the theoretical inconsistency of our  description.

 We cannot avoid the main assumption that the non-perturbative $b$ dependence is absorbed in the impact parameter behaviour of the saturation scale.  However we are planning to improve the matching procedure given by \eq{MC},
 assuming the geometric scaling behaviour of the scattering amplitude as it stems from the form of $z$ dependence  at large $z$ of the scattering amplitude
 found in this paper.

  \section{Acknowledgements}
   We thank our colleagues at Tel Aviv university and UTFSM for
 encouraging discussions. Our special thanks go to Asher Gotsman, 
 Alex Kovner and Misha Lublinsky for elucidating discussions on the
 subject of this paper.
   This research was supported by the BSF grant   2012124, by    Proyecto Basal FB 0821(Chile) ,  Fondecyt (Chile) grants 1130549 and  1140842,   CONICYT grant PIA ACT1406  and by  DGIP/USM grant 11.15.41.  
 \appendix
\section{Resumed kernel of the NLO BFKL equation }
 For completeness of presentation we collect in this appendix all formulae for the NLO kernel of the BFKL equation, \cite{BFKLNLO} resumed accordingly to the procedure, suggested in Ref. \cite{SALAM}.
 
 \bea
 \chi^{NLO}\Lb f \Rb\,&=&\, -\frac{1}{4} \Bigg(  2 b \left(\chi'(f)+\chi(f)^2\right)+\chi''(f)-\left(\frac{67}{9}-\frac{\pi ^2}{3}-\frac{10}{9}\right) \chi(f)\\
 &+&\frac{\pi ^2 \cos (\pi  f) \left(\frac{(3 f (1-f) +2) \left(\frac{N_f}{N_c^3}+1\right)}{(3-2 f) (2 f+1)}+3\right)}{(1-2 f) \sin ^2(\pi  f)}  +  4 \phi(f)-\frac{\pi ^3}{\sin (\pi  f)}-6 \zeta (3)\Bigg)\nn\\
 & - &\h \chi(f) \chi'(f) + \frac{\chi(f)}{(1 - f)^2 }   \\
   \phi(f)\,&=&\,\sum _{n=1}^\infty (-1)^n \left(\frac{\psi (-f+n+2)-\psi (1)}{(-f+n+1)^2}+\frac{\psi (f+n+1)-\psi (1)}{(f+n)^2}\right)  \\   
A_{GG}\Lb \omega\Rb &=&   b - \frac{1}{\omega +1}+\frac{1}{\omega +2}-\frac{1}{\omega +3} - \left(\psi (\omega +2)-\psi (1)\right) \nn\\
A_{QG}\Lb \omega\Rb &=& \frac{N_f}{N_c + 2}\,\left(-\frac{2}{\omega +2}+\frac{2}{\omega +3}+\frac{1}{\omega +1}\right) \nn\\
AA\Lb \omega\Rb \,&=&A_{GG}(\omega )+\frac{C_f}{N_c}\,A_{QG}(\omega )
\eea 

\bea \label{A5}
&b \,=\,\frac{11 N_c \,-\,2 N_f}{12\,N_c};~~~~C_F\,=\,\frac{N^2_c -1}{2\,N_c};~~~F = \frac{N_f}{6 \,N_c}\Lb \frac{5}{3} + \frac{13}{6  N^2_c}\Rb;&\nn\\
&\bas\Lb p^2\Rb\,\,=\,\,\frac{1}{b \,\ln\Lb p^2/\Lambda_{QCD}^2\Rb}\,\,=\,\,\frac{\bas(\mu)}{1 \,+\,b\,\bas(\mu)\,\ln\Lb p^2/\mu^2\Rb}&
\eea

In Ref. \cite{KMRS} a very elegant form of $\chi_1\Lb \omega,\gamma\Rb$i was suggested  which coincides  with \eq{SM4} to within  $7\%$. The equation for $\omega$  takes the form
\beq \label{KMRSOM}
\omega \,=\,\bas\Lb 1 - \omega\Rb \Lb \frac{1}{f}  + \frac{1}{1 - f + \omega}\,+\,\underbrace{\Lb 2 \psi(1) - \psi\Lb 2 - f\Rb -  \psi\Lb 1 + f\Rb\Rb}_{\mbox{ high twist contributions}}\Rb
\eeq
One can see that $\gamma(\omega) \to 0$ when $\omega \to 1$ as follows from energy conservation.

\begin{figure}[h]
    \centering
  \leavevmode
      \includegraphics[width=12cm]{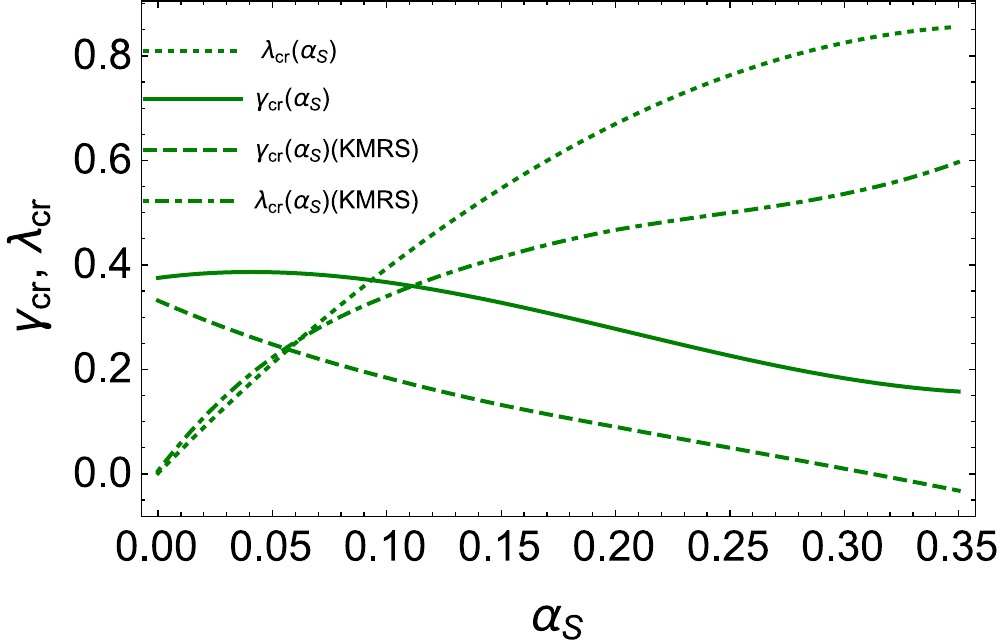}  
      \caption{$\lambda(\gamma_{cr})$ and $\gamma_{cr}$ versus $\as$ for the full NLO kernel of \eq{SM4} and for the kernel of \eq{KMRSOM}.}
\label{lagakmrs}
   \end{figure}

In \fig{lagakmrs} we plot the values of $\lambda_{cr}$ and $\gamma_{cr}$ for the full kernel of \eq{SM4} and for the simplified kernel of \eq{KMRSOM} suggested in Ref. \cite{KMRS}. One can see that in spite of the fact that the simplified kernel coincides with the full one to within $7 \%$,  the difference in $\lambda_{cr}$ and in $\gamma_{cr}$ are much larger.


    \section{Calculation of integrals for the solution in the saturation region}
    
In this appendix we take the integral of \eq{BKLT}, which has the form
\beq
\label{ecuacionprincipal}
\frac{d S_{12}}{d Y}=-\frac{\bas}{2\pi}K(Q_{s , }x_{12})\,S_{12}
\end{equation}
where
\bea
\label{kernel}
&&K(Q_{s}, x_{12})=\\
&&\int d^{2}x_{3}\frac{x^{2}_{12}}{x^{2}_{13}x^{2}_{23}}\left[ 1+\bas b  \left(\ln(\mu^{2}x^{2}_{12}) -\frac{x^{2}_{13}-x^{2}_{23}}{x^{2}_{12}}\ln\frac{x^{2}_{13}}{x^{2}_{23}} \right) +\bas\left( \frac{67}{36}-\frac{\pi^{2}}{12}-\frac{5}{18}\frac{N_{f}}{N_{c}} -\frac{1}{2}\ln\frac{x_{13}^{2}}{x_{12}^{2}}\ln\frac{x_{23}^{2}}{x_{12}^{2}} \right)  \right]\nn
\eea
Introducing the following notations
\begin{equation*}
\begin{array}{rcl}
I_{1}&=&\displaystyle{ \int d^{2}x_{3}\frac{x^{2}_{12}}{x^{2}_{13}x^{2}_{23}}\left[ 1+\bas b  \ln(\mu^{2} x^{2}_{12})+\bas \left( \frac{67}{36}-\frac{\pi^{2}}{12}-\frac{5}{18}\frac{N_{f}}{N_{c}}\right)\right]}\\
I_{2}&=&\displaystyle{ \bas b \int d^{2}x_{3} \frac{x^{2}_{12}}{x^{2}_{13}x^{2}_{23}}\left[ \frac{x^{2}_{23}-x^{2}_{13}}{x^{2}_{12}}\ln\left( \frac{x_{13}}{x_{23}}  \right) \right] }\\
I_{3}&=&\displaystyle{ -\frac{\bas}{2}\int d^{2}z_{3}\frac{x^{2}_{12}}{x^{2}_{13}x^{2}_{23}}\ln\frac{x_{13}^{2}}{x_{12}^{2}}\ln\frac{x_{23}^{2}}{x_{12}^{2}} }
\end{array}
\end{equation*}
we  have
$$ K(Q_{s}z_{12})=I_{1}+I_{2}+I_{3}  $$
Using the symmetry of the integrand with respect to $x_{13} \leftrightarrow x_{32}$ we obtain
\begin{equation}
\label{forma1}
\begin{array}{rcl}
I_{2} &=&\displaystyle{ \al\beta\int d^{2}x_{3} \frac{1}{x^{2}_{13}}\ln\left( \frac{x_{13}}{x_{23}}  \right)-\bas b \int d^{2}x_{3} \frac{1}{x^{2}_{23}}\ln\left( \frac{x_{13}}{x_{23}}  \right) }\\
      &=&\displaystyle{ \al\beta\int d^{2}x_{3} \frac{1}{x^{2}_{13}}\ln\left( \frac{x_{13}}{x_{23}}  \right)+\bas b \int d^{2}x_{3} \frac{1}{x^{2}_{23}}\ln\left( \frac{x_{23}}{x_{13}}  \right) }\\
      &=&\displaystyle{ \al\beta\int d^{2}x_{13} \frac{1}{x^{2}_{13}}\ln\left( \frac{x_{13}}{x_{23}}  \right)+\bas b \int d^{2}x_{23} \frac{1}{x^{2}_{23}}\ln\left( \frac{x_{23}}{x_{13}}  \right) }
      = \displaystyle{ 2\al\beta\int d^{2}x_{13} \frac{1}{x^{2}_{13}}\ln\left( \frac{x_{13}}{x_{23}}  \right)}
\end{array}
\end{equation}
  $I_i$ in polar coordinates take the form
\begin{equation}
\label{polar}
\begin{array}{rcl}
I_{1}&=&\displaystyle{\frac{1}{2} \int_{r_{0}^{2}}^{r_{1}^{2}}\frac{dr^{2}}{r^{2}}\int_{0}^{2\pi} \frac{1}{1+r^{2}-2r\cos\theta}d\theta\left[ 1+\bas b\ln(\mu^{2}x^{2}_{12})+\bas \left( \frac{67}{36}-\frac{\pi^{2}}{12}-\frac{5}{18}\frac{N_{f}}{N_{c}}\right)\right]};\\
\\
I_{2}&=&\displaystyle{\bas b \int_{r_{0}^{2}}^{r_{1}^{2}}\frac{dr^{2}}{r^{2}}\int_{0}^{2\pi} \ln\left(\frac{r^{2}}{1+r^{2}-2r\cos\theta}  \right)d\theta};\\
I_{3}&=&\displaystyle{-\frac{\bas}{4}\int_{r_{0}^{2}}^{r_{1}^{2}}\frac{dr^{2}}{r^{2}}\int_{0}^{2\pi} \frac{1}{1+r^{2}-2r\cos\theta}\ln(r^{2})\ln (1+r^{2}-2r\cos\theta)  d\theta}
\end{array}
\end{equation}
where $ r^2 = x^2_{13}/x^2_{12}$,  
$ r_{0}^{2}=1/Q_{s}^{2}z_{12}^{2}$ and $r^2_{1}=1- 1/Q_{s}^{2}z_{12}^{2}$.
 We use the following representations to take integral over
 the angle (see Ref. \cite{RY} formulae {\bf 1.448, 1.511,3.613}):
\begin{equation}
\label{representaciones}
\begin{array}{rcl}
\displaystyle{\int_{0}^{2\pi}\frac{\cos(n\theta)}{1+r^{2}-2r\cos\theta}d\theta }&=&\displaystyle{\frac{\Gamma(n+1)2\pi}{\Gamma(1)n!}r^{n}\ _{2}F_{1}(1,n+1;n+1,r^{2})=  2\pi\frac{r^{n}}{1-r^{2}} }\\
\ln(1+r^{2}-2r\cos\theta)&=&-2\displaystyle{\sum_{n=1}^{\infty}\frac{\cos(n\theta)}{n}r^{n}};~~~~~~~~
\ln(1-r^{2}) = \displaystyle{ -\sum_{n=1}^{\infty}\frac{r^{2n}}{n}}
\end{array}
\end{equation}
Using  \eqref{representaciones} for  $Q_{s}^{2}z_{12}^{2}\gg 1$ we obtain
\begin{equation}
\label{cadatermino}
\begin{array}{rcl}
I_{1}&=&\displaystyle{ \pi \int_{r_{0}^{2}}^{r_{1}^{2}}dr^{2}\frac{1}{r^{2}(1-r^{2})}\left[ 1+\bas b \ln(\mu^{2}x_{12}^{2})+\bas\left( \frac{67}{36}-\frac{\pi^{2}}{12}-5\frac{N_{f}}{N_{c}} \right) \right] }\\
\\
     &=&\displaystyle{2\pi \left[ 1+\bas b  \ln(\mu^{2}x_{12}^{2})+\bas\left( \frac{67}{36}-\frac{\pi^{2}}{12}-5\frac{N_{f}}{N_{c}} \right) \right]\ln(Q_{s}^{2}x_{12}^{2}) }
\\
I_{2}&=&\displaystyle{2\pi\bas b \int_{r_{0}^{2}}^{r_{1}^{2}}dr^{2}\frac{\ln r^{2} }{r^{2}}  d\theta}\,=\,\displaystyle{ -\pi\bas b  \ln^{2}Q_{s}^{2}x_{12}^{2}  }
\\
I_{3}&=&\displaystyle{-\frac{\bas}{4}\int_{r_{0}^{2}}^{r_{1}^{2}}dr^{2}\frac{\ln(r^{2})}{r^{2}}\int_{0}^{2\pi} \frac{1}{1+r^{2}-2r\cos\theta}\ln (1+r^{2}-2r\cos\theta)  d\theta  }\\
    \\
    &=&\displaystyle{-\frac{\bas}{4}\int_{r_{0}^{2}}^{r_{1}^{2}}dr^{2}\frac{\ln(r^{2})}{r^{2}}(-2)\sum_{n=1}^{\infty}\frac{r^{n}}{n}\int_{0}^{2\pi} \frac{\cos\theta}{1+r^{2}-2r\cos\theta} d\theta  }\\
    \\
    &=&\displaystyle{-\frac{\bas}{4}\int_{r_{0}^{2}}^{r_{1}^{2}}dr^{2}\frac{\ln(r^{2})}{r^{2}}(-2)\sum_{n=1}^{\infty}\frac{r^{n}}{n}\left(\frac{2\pi r^{n}}{1-r^{2}}\right)d\theta  }\\
    \\
    &=&\displaystyle{-\frac{2\pi\bas}{2}\int_{r_{0}^{2}}^{r_{1}^{2}}dr^{2}\frac{\ln(r^{2})\ln(1-r^{2}) }{r^{2}(1-r^{2})}d\theta  }\,=\,\displaystyle{-\frac{2\pi}{16}\zeta(3)   }
\end{array}
\end{equation}

Hence, we obtain the  following expression
\bea
\label{aprox}
-\frac{\bas}{2\pi}K(Q_{s}z_{12})&=&\\
& & \displaystyle{-\bas\left[ 1+\bas b  \ln(\mu^{2}z_{12}^{2})+\bas\left( \frac{67}{36}-\frac{\pi^{2}}{12}-5\frac{N_{f}}{N_{c}} \right) \right]\ln(Q_{s}^{2}z_{12}^{2}) +\frac{\bas^{2}\beta}{2} \ln^{2}Q_{s}^{2}z_{12}^{2} +\frac{\bas\zeta(3) }{16} }\nn
\eea
which we have used in section 2.4.

\end{document}